%% file: RR-aut.tex
\documentclass[a4paper]{article}
\usepackage{RR}
\usepackage{hyperref}
\usepackage{amsthm}
\usepackage{amsmath,amssymb,epsfig}
\usepackage[T1]{fontenc}
\usepackage{natbib}
\usepackage{graphicx}
\graphicspath{{figures/}}

\RRdate{Février 2007}

\RRauthor{Gr\'egory Batt%
  \thanks[inria]{INRIA Rh\^{o}ne-Alpes, 655 avenue de l'Europe, Montbonnot, 38334 Saint Ismier Cedex, France}%
  \thanks[bu]{Center for Information and Systems Engineering and Center for BioDynamics, Boston University, 15 Saint Mary's Street, Brookline, MA 02446, USA}
  \and Delphine Ropers\thanksref{inria}
  \and Hidde de Jong\thanksref{inria}
  \and Michel Page\thanksref{inria}%
  \thanks{Ecole Sup\'erieure des Affaires, Universit\'e Pierre Mend\`es France, BP 47, 38040 Grenoble Cedex 9, France}
  \and Johannes Geiselmann%
  \thanks{Laboratoire Adaptation et Pathog\'enie des Microorganismes (CNRS UMR 5163), Universit\'e Joseph Fourier, B\^{a}timent Jean Roget, Facult\'{e} de M\'{e}decine-Pharmacie, Domaine de la Merci, 38700 La Tronche, France}
}
\authorhead{G. Batt \textit{et al.}}
\RRetitle{Symbolic Reachability Analysis of Genetic Regulatory Networks using Qualitative Abstractions%
\footnote{This technical report is a revised and extended version of our previous work in~\cite{gr259}.}}
\titlehead{Symbolic Reachability Analysis of Genetic Regulatory Networks using Qualitative Abstractions}
\RRtitle{Analyse symbolique d'atteignabilit\'e de r\'eseaux de r\'egulation g\'enique par abstractions qualitatives}

\RRabstract{The switch-like character of gene regulation has motivated the use of hybrid, discrete-continuous models of genetic regulatory networks. While powerful techniques for the analysis, verification, and control of hybrid systems have been developed, the specificities of the biological application domain pose a number of challenges, notably the absence of quantitative information on parameter values and the size and complexity of networks of biological interest. We introduce a method for the analysis of reachability properties of genetic regulatory networks that is based on a class of discontinuous piecewise-affine (PA) differential equations well-adapted to the above constraints. More specifically, we introduce a hyperrectangular partition of the state space that forms the basis for a discrete abstraction preserving the sign of the derivatives of the state variables. The resulting discrete transition system provides a conservative approximation of the qualitative dynamics of the network and can be efficiently computed in a symbolic manner from inequality constraints on the parameters. The method has been implemented in the computer tool Genetic Network Analyzer (GNA), which has been applied to the analysis of a regulatory system whose functioning is not well-understood by biologists, the nutritional stress response in the bacterium \textit{Escherichia coli}.}
\RRkeyword{Piecewise-affine differential equations, qualitative analysis, discrete abstraction, hybrid systems, genetic regulatory networks, systems biology, nutritional stress response, \textit{Escherichia coli}}
\RRresume{Le fonctionnement binaire \textit{allumé/éteint} des gènes a motivé l'utilisation de modèles hybrides pour représenter les réseaux de régulations géniques.
Bien que des techni\-ques efficaces aient été développées pour l'analyse, la vérification et le contrôle des systèmes hybrides,
les spécificités des systèmes biologiques considérés, en particulier  l'absence d'infor\-mations quantitatives sur les valeurs des paramètres et la taille et la complexité des réseaux d'intérêt biologique,
posent un certain nombre de problèmes nouveaux.
Nous proposons une méthode pour l'analyse des propriétés d'atteignabilité des réseaux de régulations géniques basée
sur une classe de modèles bien adaptés aux contraintes mentionnées ci-dessus, utilisant des équations différentielles discontinues et affines par morceaux.
Plus précisément, nous introduisons une partition hyperrectangulaire de l'espace d'état, servant à définir une abstrac\-tion discrète qui préserve les signes des dérivées des variables d'état.
Le système de transitions discret ainsi défini est une approximation conservative de la dynamique du réseau et peut être calculé symboliquement de façon efficace à partir de contraintes d'inégalité sur les paramètres.
Cette méthode a été implémentée dans l'outil Genetic Network Analyzer (GNA), et utilisée pour l'analyse d'un système biologique encore mal compris,
celui de la réponse au stress nutritionnel chez la bactérie \textit{Escherichia coli}.}
\RRmotcle{Equations différentielles affines par morceaux, analyse qualitative, abstraction discrète, systèmes hybrides, réseaux de régulations géniques, biologie des systèmes, réponse au stress nutritionnel, \textit{Escherichia coli}}

\RRprojet{HELIX}
\RRtheme{\THBio}
\URRhoneAlpes

\newtheorem{df}{Definition}
\newtheorem{tm}{Theorem}

\newtheorem{lem}{Lemma}
\newtheorem{pr}{Proposition}
\newcommand{\diag}{\mbox{\rm diag}}
\newcommand{\co}{\overline{\mathit{co}}}
\newcommand{\rect}{\overline{\mathit{rect}}}
\newcommand{\maX}{\mathit{max}}
\newcommand{\sign}{\mathit{sign}}
\newcommand{\Dsign}{\mathit{Dsign}}
\newcommand{\mode}{\mathit{mode}}
\newcommand{\flow}{\mathit{flow}}
\newcommand{\supp}{\mathit{supp}}

\newcommand{\cya}{\mathit{cya}}
\newcommand{\crp}{\mathit{crp}}
\newcommand{\topA}{\mathit{topA}}
\newcommand{\rrn}{\mathit{rrn}}
\newcommand{\gyrAB}{\mathit{gyrAB}}
\newcommand{\fis}{\mathit{fis}}
\newcommand{\signal}{\mathit{signal}}
\newcommand{\op}[1]{\mathbf{#1}}

\begin{document}

\makeRR


\section{Introduction}
\label{sec:intro}

The functioning and development of living organisms is controlled on the molecular level by networks of genes, proteins, small molecules, and their mutual interactions, so-called \textit{genetic regulatory networks}. The dynamics of these networks is hybrid in nature, in the sense that the continuous evolution of the concentration of proteins and other molecules is punctuated by discrete changes in the activity of genes coding for the proteins. The switch-like character of the dynamics of genetic regulatory networks has attracted much attention from researchers on hybrid systems (\textit{e.g.}, \cite{HdJ2527,HdJ2536,HdJ2039,HdJ2354,HdJ2476}).

While powerful techniques for the analysis, verification, and control of hybrid systems have been developed (see \cite{HdJ2541,HdJ2374} for reviews), the specificities of the biological application domain pose a number of challenges \cite{HdJ1636,HdJ2528}. First, most genetic regulatory networks of interest consist of a large number of genes that are involved in complex, interlocking feedback loops. Second, the data available on both the structure and the dynamics of the networks is currently essentially qualitative in nature, meaning that numerical values for concentration variables and kinetic parameters describing the interactions are generally absent. The above characteristics require hybrid-system methods and tools to be upscalable and capable of dealing with qualitative information.

In this paper, we will show that a class of piecewise-affine differential equation (PADE) models, originally introduced by Glass and Kauffman in the seventies \cite{HdJ1030}, is particularly suitable for dealing with the above challenges. The properties of these PADE models have been well-studied in mathematical biology \cite{HdJ2474,HdJ2516,HdJ2125,HdJ1619,HdJ1553,HdJ2515,HdJ1029,HdJ1195,HdJ1899,HdJ1032,HdJ2488,HdJ1038}. The variables in the PADE models are the concentrations of the proteins encoded by the genes, while step functions account for the discrete changes in gene activity occasioned by the regulatory interactions. On a formal level, the PADE models are related to a class of asynchronous logical models proposed by Thomas and colleagues \cite{HdJ1039,HdJ797}. PADE models and their logical relatives have been used for the study of a number of prokaryotic and eukaryotic regulatory networks (see \cite{HdJ2426} for a review and references).

The particular form of the PADE models allows the continuous state space to be partitioned into hyperrectangular regions with equivalent qualitative dynamics, preserving the sign pattern of the time derivatives of the solutions. We exploit this partition by associating to each PADE model a continuous transition system having equivalent reachability properties. By means of a \textit{discrete} or \textit{qualitative abstraction} \cite{HdJ2540,HdJ2529,HdJ2205,HdJ2370,HdJ2530,HdJ1908}, the continuous transition system is turned into a discrete transition system providing a compact and qualitative description of the dynamics of the continuous system. Formally, there exists a simulation relation between the continuous and discrete transition systems, that is, the latter is a conservative approximation of the former. We show that the discrete transition system is invariant over large ranges of parameter values and can be computed in a symbolic manner from inequality constraints. Moreover, it is possible to design tailored algorithms for computing the discrete transition system, which scale up to large and complex genetic regulatory networks.

The paper continues our previous work on the qualitative analysis of the dynamics of genetic regulatory networks by means of PADE models \cite{HdJ2039,HdJ2125}. The main novelty of this contribution, a preliminary version of which was presented in \cite{HdJ2378}, is the use of a more fine-grained discrete abstraction, preserving the derivative sign pattern. This is an important feature for the experimental validation of models of genetic regulatory networks, since measurements of gene expression often result in observations of changes in the sign of derivatives. Notwithstanding this increase in precision, the refinement does not threaten the applicability of the approach to large and complex networks. This is illustrated by the analysis of a network that is not yet well understood by biologists, the network controlling the carbon starvation response of the bacterium \textit{Escherichia coli}. The application of the method has led to novel insights into how the adaptation of cell growth to nutritional stress emerges from the network of molecular interactions.

The paper is organized as follows. In Sections~\ref{sec:PADEmodels} and \ref{sec:mathanal} we specify the PADE models of genetic regulatory networks in detail and discuss their mathematical properties, paying special attention to complications arising from the discontinuities in the righthand-side of the differential equations. In Section~\ref{sec:abstraction} we define a qualitative abstraction of the dynamics of PA systems, based on a hyperrectangular partition of the state space. Section~\ref{sec:computation} introduces rules to actually compute the discrete transition system induced by the qualitative abstraction and the implementation of the rules in a computer tool called GNA. In Section~\ref{sec:application} we illustrate the application of the method to the analysis of the \textit{E. coli} network. In the final section we present our conclusions and discuss the results in the context of related work.

\section{PADE models of genetic regulatory networks}
\label{sec:PADEmodels}

The dynamics of genetic regulatory networks can be described by a class of piecewise-affine differential equations (PADE) models of the following general form \cite{HdJ1030, HdJ1032}:
\begin{equation}
\dot{x} = h(x)= f(x)-g(x)\, x, \quad x \in \Omega \setminus \Theta,
\label{eq:kinbasic}
\end{equation}
where $x=(x_{1},\ldots ,x_{n})'$ is a vector of cellular protein concentrations, $f=(f_{1},\ldots ,f_{n})'$, $g=\diag(g_{1},\ldots ,g_{n})$, $\Omega\subset \mathbb{R}^n_{\geq 0}$ is a bounded $n$-dimensional state space region, and $\Theta$ a zero-measure subset of $\Omega$ (see below). The rate of change of each protein concentration $x_{i}$, $i \in \{ 1, \ldots , n\}$, is thus defined as the difference of the rate of synthesis $f_{i}(x)$ and the rate of degradation $g_{i}(x)\, x_{i}$ of the protein.

The function $f_{i}:\ \Omega \setminus \Theta \to \mathbb{R}_{\geq 0}$ expresses how the rate of synthesis  of the protein encoded by gene $i$ depends on the concentrations $x$ of the  proteins in the cell. More specifically, the function $f_{i}$ is defined as
\begin{equation}
f_{i}(x) = \sum_{l\in L_i} \kappa_{i}^l\, b_{i}^l(x),
\label{eq:prodterm}
\end{equation}
where $\kappa_{i}^l>0$ is a rate parameter, $b_{i}^l:\ \Omega \setminus \Theta \to \{ 0,1\}$ a piecewise-constant \emph{regulation function}, and $L_i$ a  possibly empty set of indices of regulation functions. The function $g_{i}$ expresses the regulation of protein degradation. It is defined analogously to $f_{i}$, except that we demand that $g_{i}$ is strictly positive. In addition, in order to formally distinguish degradation rate parameters from synthesis rate parameters, we will denote the former by $\gamma$ instead of $\kappa$. Notice that with the above definitions, $h$ is a \textit{piecewise-affine (PA)} vector-valued function.

A regulation function  $b_{i}^l$ describes the conditions under which the protein encoded by gene $i$ is synthesized (degraded) at a rate $\kappa_{i}^l$ ($\gamma_{i}^l\, x_i$). It is defined in terms of step functions and is the arithmetic equivalent of a Boolean function expressing the logic of gene regulation \cite{HdJ1030,HdJ1039}. More precisely, the conditions for synthesis and degradation are expressed using the step functions $s^+, s^-$:
\begin{equation}
  s^+(x_j, \theta_j)= \left\{
    \begin{array}{r@{~\mbox{if}~}l}
      1,& x_j> \theta_j,\\
      0,& x_j< \theta_j,
    \end{array} \right.
    \hspace*{.2cm} s^-(x_j, \theta_j)= 1-s^+(x_j, \theta_j),
\end{equation}
where $x_j$ is an element of the state vector $x$ and $\theta_j$ a constant denoting a threshold concentration. Notice that the step functions are not defined at the thresholds.

\begin{figure*}[htp]
\begin{minipage}[b]{14cm}
\begin{minipage}[b]{7.5cm}
\begin{center}
\scalebox{.50}{\input{figures/exnetwork.pstex_t}}\\
\mbox{(a)}
\end{center}
\end{minipage}\hfill
\begin{minipage}[b]{6.5cm}
\begin{center}
\begin{align*}
\dot{x}_{a} & = \kappa_{a} \, s^{-}(x_{a},\theta_{a}^2)\, s^{-}(x_{b},\theta_{b})
              - \gamma_{a}\, x_{a}, \\
\dot{x}_{b} & = \kappa_{b} \,  s^{-}(x_{a},\theta_{a}^1)
              - \gamma_{b}\, x_{b}.
\end{align*}
\mbox{(b)}
\end{center}
\end{minipage}
\end{minipage}
\caption{\label{fig:example} (a) Example of a genetic regulatory network of two genes (\textit{a} and \textit{b}), each coding for a regulatory protein (A and B). See Figure~\ref{fig:reseau} for the legend. (b) PADE model corresponding to the network in (a).}
\end{figure*}
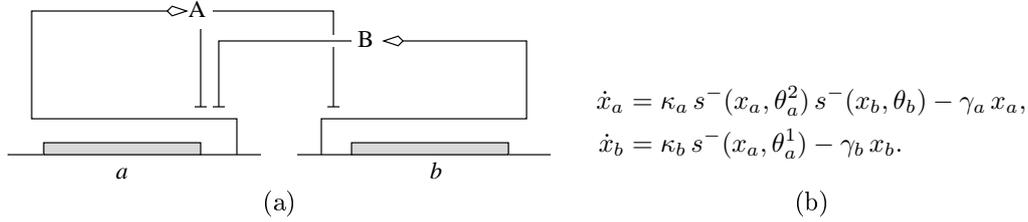

Figure~\ref{fig:example}(a) gives an example of a simple genetic regulatory network consisting of two genes, \textit{a} and \textit{b}. When a gene (\textit{a} or \textit{b}) is expressed, the corresponding protein (A or B) is synthesized at a specified rate ($\kappa_a$ or $\kappa_b$). Proteins A and B regulate the expression of genes \textit{a} and \textit{b}. More specifically, protein B inhibits the expression of gene \textit{a}, above a certain threshold concentration $\theta_b$, while protein A inhibits the expression of gene \textit{b} above a threshold concentration $\theta_a^1$, and the expression of its own gene above a second, higher threshold concentration $\theta_a^2$. The degradation of the proteins is not regulated and proportional to the concentration of the proteins (with degradation parameters $\gamma_a$ or $\gamma_b$). The PADE model of this network is shown in Figure~\ref{fig:example}(b).

The use of step functions $s^{\pm}(x_j, \theta_j)$ in (\ref{eq:kinbasic}) gives rise to mathematical complications, because the step functions are undefined and discontinuous at $x_j=\theta_j$. Therefore, $h$ is undefined and may be discontinuous on the threshold hyperplanes $\Theta = \bigcup_{i \in \{ 1, \ldots ,n\}, l_i \in \{ 1, \ldots ,p_i\}} \{ x \in \Omega \mid  x_i = \theta_i^{l_i}\}$, where $p_i$ denotes the number of thresholds for the protein encoded by gene $i$. In order to deal with this problem, we can follow an approach originally proposed by Filippov \cite{HdJ1661} and widely used in control theory. It consists in extending the differential \textit{equation} $\dot{x} = h(x)$, $x \in \Omega \setminus \Theta$, to the differential \textit{inclusion}
\begin{equation}
\label{eq:kininclusion_f}
\dot{x}\in K(x), \mbox{ with }  K(x)= \co(\{ \lim_{y\to x,\, y \not \in \Theta} h(y)\}),\; x \in \Omega,
\end{equation}
where $\co(P)$ denotes the smallest closed convex set containing the set $P$, and \linebreak $\{ \lim_{y\to x,\, y\not \in \Theta} h(y)\}$ the set of all limit values of $h(y)$, for $y\not \in \Theta$ and $y\to x$. This approach has been applied in the context of genetic regulatory network modeling by Gouz\'{e} and Sari \cite{HdJ1899}.

In practice, $K(x)$ may be difficult to compute because the smallest closed convex set can be a complex polyhedron in $\Omega$. We therefore employ an alternative extension of the differential equation:
 \begin{equation}
    \label{eq:kininclusion}
\dot{x}\in H(x), \mbox{ with }  H(x)= \rect(\{ \lim_{y\to x,\, y\not \in \Theta} h(y)\}),\, x \in \Omega,
  \end{equation}
where $\rect(P)$ denotes the smallest closed \textit{hyperrectangular} set containing the set $P$. The advantage of using $\rect$ is that we can rewrite $H(x)$ as a system of differential inclusions $\dot{x}_i \in H_i(x),\ i\in \{1, \ldots ,n\}$. Notice that $H(x)$ is an overapproximation of $K(x)$, for all $x\in \Omega$.

Formally, we define the \textit{PA system} $\Sigma$ as the triple $(\Omega, \Theta, H)$, that is, the set-valued function $H$ given by (\ref{eq:kininclusion}), and defined on the $n$-dimensional state space $\Omega$, with $\Theta$ the union of the threshold hyperplanes. A \emph{solution} of the PA system $\Sigma$ on a time interval $I$ is a solution of the differential inclusion (\ref{eq:kininclusion}) on $I$, that is, an absolutely-continuous vector-valued function $\xi(t)$ such that $\dot{\xi}(t)\in H(\xi(t))$ almost everywhere on $I$. In particular, the derivative of $\xi(t)$ may not exist, and therefore $\dot{\xi}(t)\in H(\xi(t))$ may not hold, if $\xi$ reaches or leaves $\Theta$ at $t$.

For all $x_0\in \Omega$ and $\tau \in \mathbb{R}_{>0}\cup \{\infty\}$, $\Xi_\Sigma(x_0, \tau)$ denotes the set of solutions $\xi(t)$ of the PA system $\Sigma$, for the initial condition $\xi(0)=x_0$, and $t \in [0,\tau]$. The existence of at least one solution $\xi$ on some time interval $[0, \tau]$, $\tau>0$, with initial condition $\xi(0)=x_0$  is guaranteed for all $x_0$ in $\Omega$ \cite{HdJ1661}. However, there is, in general, not a unique solution. The set $\Xi_\Sigma= \bigcup_{x_0 \in \Omega, \tau> 0} \Xi_\Sigma(x_0, \tau)$ is the set of all solutions, on a finite or infinite time interval, of the PA system $\Sigma$. We restrict our analysis to the solutions in $\Xi_\Sigma$ that reach and leave a threshold hyperplane finitely-many times. The dynamics of $\Sigma$ is thus defined by the set of solutions $\Xi_\Sigma$.

\section{Mathematical analysis of PA systems}
\label{sec:mathanal}

\subsection{Mode domains}
\label{sec:mathanal,mode}

The dynamical properties of the solutions of $\Sigma$ can be analyzed in the $n$-dimensional state space hyperrectangle $\Omega=\Omega_{1}\times \ldots \times \Omega_{n}$, where $\Omega_{i} = \{x_i \in \mathbb{R} \mid 0 \leq x_i \leq \maX_i\}$ and $\maX_{i}$ denotes a maximum concentration for each protein, $i\in \{ 1, \ldots ,n\}$. In particular, we set $\maX_i > \max_{x \in \Omega \setminus \Theta} f_{i}(x) / g_{i}(x)$. It is not difficult to show that under this condition $\Omega$ is a positively invariant set.

For subsequent use, we introduce the notion of \textit{hyperrectangular partition} of a set $R = R_{1}\times \ldots \times R_{n} \subseteq \Omega$. The partition is induced by sets of hyperplanes orthogonal to one of the axes $x_i$, $i\in \{1, \ldots ,n\}$. More precisely, the hyperplanes orthogonal to the $x_i$-axis are given by the finite sets $\Lambda_i \subset \Omega_{i}$, where every $\lambda \in \Lambda_i$ corresponds to a hyperplane $\{ x \in \Omega \mid x_i = \lambda\}$. The hyperrectangular partition of $R$ induced by $\Lambda = \{ \Lambda_1, \ldots ,\Lambda_n\}$ is defined as $\mathcal{P} = \mathcal{P}_1 \times \ldots \times \mathcal{P}_n$, where $\mathcal{P}_i$, $i\in \{ 1, \ldots ,n\}$, is the \textit{interval partition} of $R_i$ induced by $\Lambda_i$.  That is, $\mathcal{P}_i$ is the partition of minimal cardinality of $R_i$, such that for every $P \in \mathcal{P}_i$, either $P=\{ \lambda\}$ for some $\lambda \in \Lambda_i$, or $P$ is an interval containing no $\lambda \in \Lambda_i$. As an example, consider the interval $R_i = (r_1, r_2]$ and $\Lambda_i = \{\lambda_1,\lambda_2\}$, with $r_1 < \lambda_1 < r_2 < \lambda_2$. The interval partition of $R_i$ induced by $\Lambda_i$ equals $\mathcal{P}_i = \{ (r_1, \lambda_1), \{\lambda_1\}, (\lambda_1,r_2]\}$.

The \mbox{$(n-1)$-dimensional} threshold hyperplanes introduced in Section~\ref{sec:PADEmodels} define a hyperrectangular partition of $\Omega$.

\begin{df}[Mode domain partition]\rm
\label{df:mode_domain}
$\mathcal{M}$ is the hyperrectangular partition of $\Omega$ induced by $\{ \theta_i^{1}, \ldots, \theta_i^{p_i}\}$, $i\in \{ 1, \ldots ,n\}$. The sets $M \in \mathcal{M}$ are called \textit{mode domains}.
\end{df}
The reason for speaking of mode domains will become clearer below, when we show that the regulation of gene expression is identical in every $M \in \mathcal{M}$, and thus corresponds to a regulatory mode of the system.

\begin{figure*}[htb]
  \begin{minipage}[b]{14.2cm}
    \begin{minipage}[b]{7cm}
      \begin{center}
        \scalebox{.55}{\input{figures/exbox1_coarse.pstex_t}}\\
        \mbox{(a)}
      \end{center}
    \end{minipage}\hfill
    \begin{minipage}[b]{7cm}
      \begin{center}
        \scalebox{.55}{\input{figures/exbox2_coarse.pstex_t}}\\
        \mbox{(b)}
      \end{center}
    \end{minipage}
  \end{minipage}
  \caption{\label{fig:exnet-coarse_phase_space} (a) Mode domain partition of the state space for the model of Figure~\protect\ref{fig:example}(b).  \mbox{(b) Focal sets and vector fields} associated with the mode domains $M^{1}$ to $M^{5}$, and $M^{11}$.}
\end{figure*}
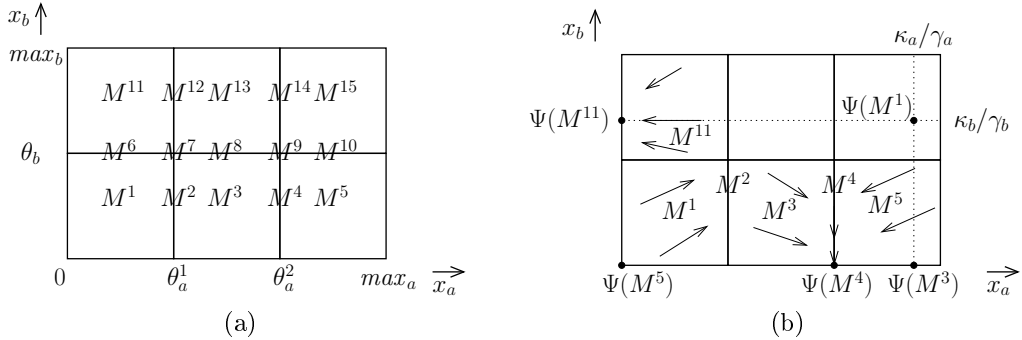

Figure~\ref{fig:exnet-coarse_phase_space}(a) shows the mode domain partition of the two-dimensional state space of the example network. We distinguish between mode domains like $M^{2}$ and $M^{7}$, which are located on (intersections of) threshold hyperplanes, and mode domains like $M^{1}$, which are not. The former domains are called \textit{singular} mode domains and the latter \textit{regular} mode domains. We denote by $\mathcal{M}_r$ and $\mathcal{M}_s$ the sets of regular and singular mode domains, respectively. Note that $\mathcal{M} = \mathcal{M}_r \cup \mathcal{M}_s$.

We next introduce some basic topological concepts. For every hyperrectangular set, $R\subseteq \Omega$, of dimension $k$, $k \in \{ 0, \ldots ,n\}$, we define the supporting hyperplane of $R$, $\supp(R) \subseteq \Omega$, as the $k$-dimensional hyperplane containing $R$. The \textit{boundary} of $R$ in $\supp(R)$, denoted by $\partial R$, is defined as the set $\overline{R}\setminus \overset{\;\circ}{R}$, where $\overline{R}$ is the closure of $R$ and $\overset{\;\circ}{R}$ its interior, both relative to $\supp(R)$. Suppose that $M$ is a singular mode domain, \textit{i.e.} $M \in \mathcal{M}_s$. Then $R(M)$ is defined as the set of regular mode domains $M'$ having $M$ in their boundary, {\it i.e.} $R(M)=\{M'\in \mathcal{M}_r \mid M\subseteq \partial M'\}$.

Using the definition of the differential inclusion (\ref{eq:kininclusion}), it can be easily shown that in a regular mode domain $M$,
\begin{equation}
H(x) = \{ \mu^{M}-\nu^{M}\, x\},\ x \in M,
\label{eq:Hregdom}
\end{equation}
where $\mu^{M}$ is a vector of (sums of) positive synthesis rate constants and $\nu^{M}=\diag(\nu^{M}_{1},\ldots ,\nu^{M}_{n})$ a diagonal matrix of (sums of) positive degradation rate constants \cite{HdJ2125}. We define the \textit{focal set}
\begin{equation}
\Psi(M) = \{ \psi(M)\},
\label{eq:Psireg}
\end{equation}
where $\psi(M)=(\nu^{M})^{-1}\, \mu^{M}$, and repeat the following standard result \cite{HdJ1030}.

\begin{lem}\rm
\label{lem:monotone-regular}
Let $M\in \mathcal{M}_{r}$. Every solution $\xi(t) \in \Xi_\Sigma$ in $M$ monotonically converges towards $\Psi(M)$.
\end{lem}

\begin{proof}
From (\ref{eq:Hregdom}) it follows that the solutions in a regular mode domain $M$ are given by $\xi(t)= \psi(M) - e^{- \nu^M  t} (\psi(M)-x_0)$,  for initial conditions $x_0\in M$. As a consequence, $\xi(t)$ monotonically converges towards $\Psi(M) = \{ \psi(M)\}$ as long as $\xi(t) \in M$.
\end{proof}

We will make the generic assumption that the focal sets $\Psi(M)$, for all $M \in \mathcal{M}_r$, are not located in the threshold hyperplanes $\Theta$. Figure~\ref{fig:exnet-coarse_phase_space}(b) shows the  focal sets of four regular mode domains ($M^{1}$, $M^{3}$, $M^{5}$ and $M^{11}$). In the case of $M^{11}$, we see that $\Psi(M^{11}) \subseteq M^{11}$, so that $\psi(M^{11})$ is an asymptotically stable equilibrium point of $\Sigma$.

In a singular mode domain, the right-hand side of the differential inclusion (\ref{eq:kininclusion}) reduces to the following set:
\begin{equation}
H(x) = \rect(\{\mu^{M'} - \nu^{M'} x \mid M'\in R(M)\}),\ x \in M.
\label{eq:Hsingdom}
\end{equation}
The \textit{focal set} associated with the domain is now defined as
\begin{equation}
\Psi(M)= \supp(M) \cap \rect(\{\psi(M') \mid M'\in R(M)\}),
\label{eq:Psising}
\end{equation}
which is generally not a single point in higher-dimensional systems. As will be shown below, two different cases can be distinguished. If $\Psi(M)$ is empty, then every solution passes through $M$ instantaneously (Lemma~\ref{lem:instantaneous-singular}). If not, then some (but not necessarily all) solutions arriving at $M$ will remain in $M$ for some time, sliding along the threshold hyperplanes containing $M$ (Lemma~\ref{lem:monotone-singular}). In the latter case, weaker monotonicity properties hold (Lemmas~\ref{lem:monotone-singular} and \ref{lem:monotone-singular-existence}).

\begin{lem}\rm
\label{lem:instantaneous-singular}
Let $M\in \mathcal{M}_{s}$. Every solution $\xi(t) \in \Xi_\Sigma$ arriving at $M$ instantaneously crosses the domain if and only if $\Psi(M) = \{ \}$.
\end{lem}

\begin{proof}
We first prove sufficiency. From  the definition of $\Psi(M)$ in a singular mode domain and $\Psi(M) = \{\}$, it follows that for some $i \in \{ 1, \ldots , n\}$ such that $M_i$ is included in a threshold hyperplane, the intersection of $\supp_i(M) = \{ \theta_i^{l_i} \}$, $l_i \in \{ 1, \ldots , p_i \}$, and $\rect(\{\psi_i(M') \mid M'\in R(M)\})$ is empty. As a consequence, either $\psi_i(M') > \theta_i^{l_i}$ for all $M'\in R(M)$, or \mbox{$\psi_i(M') < \theta_i^{l_i}$} for all $M'\in R(M)$. From (\ref{eq:Hsingdom}) and the hyperrectangular shape of $H(x)$, we obtain $H_i (x) = \rect(\{ \nu_{i}^{M'} \; (\psi_i (M') - \theta_i^{l_i}) \mid M' \in R(M) \})$, for all $x \in M$. As a consequence, either $\min H_i(x) > 0$ or $\max H_i(x) < 0$, so that $0 \not \in H_i (x)$ and no solution with $\dot{\xi}_i (t) = 0$ exists in $M$. Therefore, every solution arriving at $M$ instantaneously crosses the domain.

Necessity is proven by contraposition, by an argument similar to that given for sufficiency. From $\Psi(M) \not = \{\}$ we show that $0 \in H_i(x)$, for every $x \in M$ and $i \in \{ 1, \ldots , n\}$ such that $M_i$ is included in a threshold hyperplane. We can then define solutions $\xi(t)$ that satisfy $\dot{\xi}_i(t) =0$ on some time interval, that is, solutions that do not instantaneously cross $M$.
\end{proof}

\begin{lem}\rm
\label{lem:monotone-singular}
Let $M\in \mathcal{M}_{s}$ and $\Psi(M) \not = \{\}$. For every solution $\xi(t) \in \Xi_\Sigma$ in $M$, and every $i \in \{ 1, \ldots , n\}$ such that $M_i$ is included in a threshold hyperplane, it holds that $\dot{\xi}_i(t)=0$. For all other $i$, $\xi_i(t)$ monotonically converges towards $\Psi_i(M)$.
\end{lem}

\begin{proof}
In order to remain in $M$, $\xi(t)$ must slide along the threshold hyperplanes containing $M$. Therefore, $\dot{\xi}_i(t)=0$ for every $x \in M$ and $i \in \{ 1, \ldots , n\}$ such that $M_i$ is included in a threshold hyperplane. For all other $i$, we must have $\dot{\xi}_i(t) \in H_i(\xi(t))$ almost everywhere, by the definition of solutions of PA systems. Moreover, $H_i(\xi(t))=\rect(\{ \nu_{i}^{M'} \; (\psi_i (M') - \xi_i(t)) \mid M' \in R(M) \})$. Let $\xi_i(t) \not \in \Psi_i(M)$, and $\xi_i(t) < \psi_i(M')$ for all $M'\in R(M)$ (the case $\xi_i(t) > \psi_i(M')$ goes analogously). It follows that $\min H_i(\xi(t)) > 0$, and hence that $\dot{\xi}_i (t) > 0$, provided that $\dot{\xi}(t)\in H(\xi(t))$. We thus infer that $\xi_i(t)$ monotonically converges towards $\Psi_i(M)$.
\end{proof}

\begin{lem}\rm
\label{lem:monotone-singular-existence}
Let $M\in \mathcal{M}_{s}$ and $\Psi(M) \not = \{\}$. For every $\psi \in \Psi(M)$ and $x \in M$, there exists a solution $\xi(t) \in \Xi_\Sigma$ in $M$, with $\xi(0)=x$, such that $\xi(t)$ monotonically converges towards $\psi$.
\end{lem}

\begin{proof}
Let $M^{+}(i) = \arg \max_{M' \in R(M)} \; \psi_i(M')$ and $M^{-}(i) = \arg \min_{M' \in R(M)} \; \psi_i(M')$, for all $i \in \{ 1, \ldots , n\}$. Notice that due to the hyperrectangular shape of $\Psi(M)$, we can write $\psi_i = \alpha_i\; \psi_i(M^{-}(i)) + (1 - \alpha_i)\; \psi_i(M^{+}(i))$, for some $\alpha_i \in [0, 1]$. Moreover, after defining $\beta_i \in [0, 1]$ such that $\beta_i = \alpha_i \; \nu_i^{M^{+}(i)} / (\alpha_i \; \nu_i^{M^{+}(i)} + (1 - \alpha_i)\; \nu_i^{M^{-}(i)})$, we introduce $\mu_i$ and $\nu_i$ such that
\begin{equation}
\psi_i = \dfrac{\mu_i}{\nu_i} = \dfrac{\beta_i\; \mu_i^{M^{-}(i)} + (1 - \beta_i)\; \mu_i^{M^{+}(i)}}{\beta_i\; \nu_i^{M^{-}(i)} + (1 - \beta_i)\; \nu_i^{M^{+}(i)}}.
\label{eq:psiexp}
\end{equation}
That is, $\psi_i$ is written as the quotient of $\mu_i$ and $\nu_i$, which in turn can be defined as a convex combination of $\mu_i^{M^{-}(i)}$, $\mu_i^{M^{+}(i)}$ and $\nu_i^{M^{-}(i)}$, $\nu_i^{M^{+}(i)}$, respectively.

Now, construct a function $\xi(t)= \psi - e^{- \nu  t}\; (\psi-x)$. This yields $\dot{\xi}_i (t) = \nu_i \; e^{- \nu_i  t}\; (\psi_i-x_i) = \nu_i \; (\psi_i-\xi_i(t)) = \mu_i - \nu_i \; \xi_i(t)$, for every $i \in \{ 1, \ldots , n\}$. Following the definition of $\mu_i$ and $\nu_i$ in (\ref{eq:psiexp}), this gives
\begin{multline}
\dot{\xi}_i(t) = \beta_i\; (\mu_i^{M^{-}(i)} - \nu_i^{M^{-}(i)} \xi_i(t)) + (1 - \beta_i)\; (\mu_i^{M^{+}(i)} - \nu_i^{M^{+}(i)} \xi_i(t)).
\label{eq:xidotexp}
\end{multline}
For $\xi(t)$ to be a solution, it must satisfy $\dot{\xi}_i(t) \in H_i(\xi(t))$ almost everywhere, for every $i \in \{ 1, \ldots , n\}$. It can be directly verified that the expression for $\dot{\xi}_i(t)$ in (\ref{eq:xidotexp}), with $\beta_i \in [0,1]$, is included in $H_i(\xi(t))$ as defined by (\ref{eq:Hsingdom}). Hence, the condition is satisfied and $\xi(t)$ is a solution of $\Sigma$ in $M$. Moreover, $\xi(0)=x$ and $\xi(t)$ monotonically converges towards $\psi$.
\end{proof}

In the sequel, domains which every solution crosses instantaneously will be called \textit{instantaneous} domains, whereas domains in which at least some solutions remain for some time will be referred to as \textit{persistent}. Figure~\ref{fig:exnet-coarse_phase_space}(b) shows two singular mode domains, $M^2$ and $M^4$. $M^2$ is an instantaneous mode domain ($\Psi(M^{2})=\emptyset$), whereas $M^4$ is a persistent mode domain ($\Psi(M^{4})=\{(\theta_a^2,0)'\}$) in which solutions slide along the threshold plane. In this simple example, it is intuitively clear how to define the flow in $M^4$, given the dynamics in $M^3$ and $M^5$. The use of differential inclusions as described above makes it possible to define the flow in singular domains in a systematic and mathematically proper way.

The fact that every mode domain is associated with a unique focal set has provided the basis for a discrete abstraction criterion employed in our previous work \cite{HdJ2474,HdJ2039,HdJ2125}. However, this criterion disregards the fact that the system does not always exhibit the same qualitative dynamics in different parts of a mode domain, in the sense that the sign pattern of the derivatives of the solutions may not be unique. Consider a solution $\xi(t) \in \Xi_\Sigma$ in $M^{11}$ in Figure~\ref{fig:exnet-coarse_phase_space}(b): depending on whether $\xi_b(t)$ is larger than, equal to, or smaller than the focal concentration $\kappa_b/\gamma_b$ in $M^{11}$, $\xi_b$ will be decreasing, steady, or increasing. As a consequence, if we abstract the domain $M^{11}$ away into a single discrete state, we will not be able to unambiguously infer that solutions entering this domain from $M^{6}$ are increasing in the $x_b$-dimension. This may lead to problems when comparing predictions from the model with gene expression data, for instance the observed variation of the sign of $\dot{x}_b$. Today's experimental techniques, such as quantitative RT-PCR, reporter genes, and DNA microarrays, usually produce information on changes in the sign of the derivatives of the concentration variables.

\subsection{Flow domains}
\label{sec:mathanal,flow}

The mismatch between the description levels of the mathematical analysis and the experimental data calls for a finer partitioning of the state space, which can then provide the basis for a more adequate abstraction criterion. Along these lines, the regular and singular mode domains distinguished above are repartitioned by means of the \mbox{$(n-1)$-dimensional} hyperplanes corresponding to the focal concentrations.

\begin{df}[Flow domain partition]\rm
\label{df:flow_domain}
$\mathcal{D}^M$ is the hyperrectangular partition of a mode domain $M \in \mathcal{M}$ induced by $\{\psi_i(M) \}$, if $M$ is regular, and by $\{\psi_i(M') \mid M'\in R(M) \}$, if $M$ is singular, $i\in \{ 1, \ldots ,n\}$. $\mathcal{D}=\cup_{M\in \mathcal{M}} \mathcal{D}^M$ is a partition of $\Omega$ and the sets $D \in \mathcal{D}$ are called \textit{flow domains}.
\end{df}

\begin{figure*}[htb]
  \begin{minipage}[b]{14cm}
    \begin{minipage}[b]{7cm}
      \begin{center}
        \scalebox{.55}{\input{figures/exbox1_fine.pstex_t}}\\
        \mbox{(a)}
      \end{center}
    \end{minipage}
    \begin{minipage}[b]{7cm}
      \begin{center}
        \scalebox{.40}{\input{figures/statetrgraph_hs.pstex_t}}\\
        \hspace*{.6cm}\mbox{(b)}
      \end{center}
    \end{minipage}
    \end{minipage}\\
  \begin{minipage}{14cm}
     \begin{center}
        \begin{small}
        \begin{align*}
        & 0 < \theta_a^1 < \theta_a^2 < \kappa_a / \gamma_a < \maX_a \\
        & 0 < \theta_b < \kappa_b / \gamma_b < \maX_b
        \end{align*}
        \end{small}\\
        \mbox{(c)}
     \end{center}
  \end{minipage}
  \caption{\label{fig:partition_singular2} (a) Flow domain partition of the state space of the model in Figure~\ref{fig:example}(b). (b) State transition graph of the corresponding qualitative transition system. For the sake of clarity, self-transitions are represented by dots and transition labels are omitted. (c) Parameter inequality constraints for which the graph in (b) is invariant.}
\end{figure*}
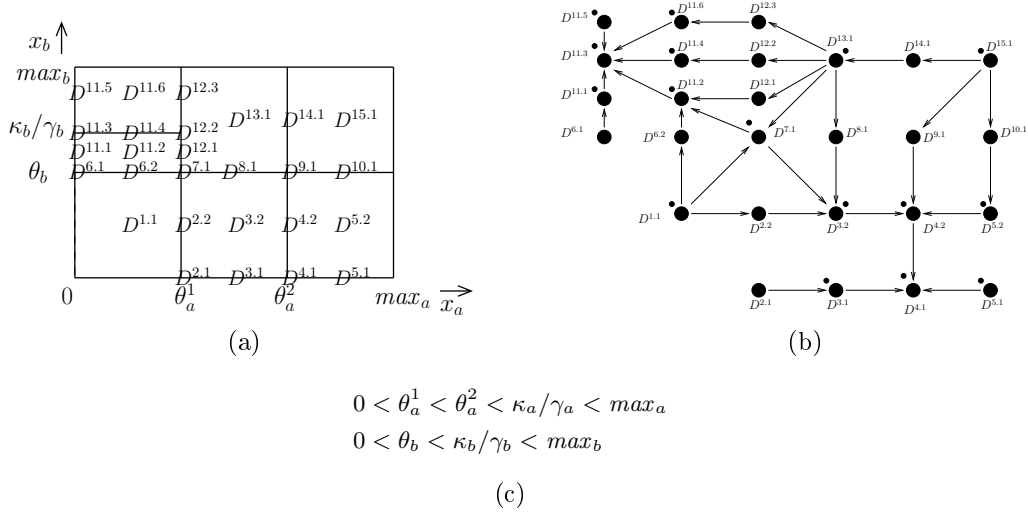

The reason for speaking about flow domains is that, as will be seen below, in every $D\in \mathcal{D}$ the flow of the system is qualitatively identical. The partitioning of the state space into 27 flow domains is illustrated for the example system in Figure~\ref{fig:partition_singular2}(a). Every flow domain is included in a single mode domain, a relation captured by the surjective function $\mode$: $\mathcal{D} \to \mathcal{M}$, defined as $\mode(D)=M$, if and only if $D\subseteq M$. Similarly, the function $\flow$: $\Omega \to \mathcal{D}$ denotes the surjective mapping that associates a point in the state space to its flow domain: $\flow(x)=D$, if and only if $x \in D$.

The repartitioning of mode domain $M^{11}$ leads to six flow domains (Figure~\ref{fig:partition_singular2}(a)). The finer partition guarantees that in every flow domain of $M^{11}$, the derivatives have a unique sign pattern. In $D^{11.2}$, for instance, the $x_a$-derivative is negative and the $x_b$-derivative is positive, whereas in $D^{11.3}$ both derivatives equal zero (in fact, $D^{11.3}$ coincides with $\psi(M^{11})$ and is an equilibrium point of the system). This property is true more generally. Consider a point $x$ in a flow domain $D\in \mathcal{D}$. We denote by $S(x) \in 2^{\{-1, 0, 1\}^n}$ the derivative sign pattern at $x$, that is, the set of derivative sign vectors of the solutions in $D$ passing through $x$. More formally,
\begin{multline}
S(x)= \{\sign(\dot{\xi}(\tau)) \mid \xi(t) \in \Xi_\Sigma \mbox{ in } D, \mbox{ and } \exists \tau  \geq 0:\ \xi(\tau)=x \mbox{ and }\dot{\xi}(\tau)\in H(\xi(\tau))\}.
\label{eq:S}
\end{multline}
This gives the following central result.

\begin{tm}[Qualitatively-identical dynamics]\rm
\label{tm:abstraction}
For all $x, y \in D$ and $D \in \mathcal{D}$, $S(x)= S(y)$.
\end{tm}

\begin{proof}
Note that due to the hyperrectangular shape of $H(x)$, $S(x)$ can be decomposed into $S_1(x) \times \ldots \times S_n(x)$, where $S_i(x)$ denotes the set of the $i$th components of the derivative sign vectors of the solutions in $D$ passing through $x$, $i \in \{ 1, \ldots ,n \}$ (idem $S(y)$). Define $M=\mode(D)$ and distinguish the cases (a)-(c).

(a) $M \in \mathcal{M}_r$. Suppose $1 \in S_i(x)$, but $1 \not \in S_i(y)$, for any $i \in \{ 1, \ldots ,n \}$ (for $0$ and $-1$ the argument is similar). By definition of $S_i(x)$ this means that $\dot{\xi}_i(\tau)>0$ for some $\xi(t) \in \Xi_\Sigma$ in $D$ and $\tau \geq 0$, such that $\xi(\tau)=x$. On the other hand, $\dot{\varphi}_i(\sigma) \leq 0$ for all $\varphi(t) \in \Xi_\Sigma$ in $D$ and $\sigma \geq 0$, such that $\varphi(\sigma)=y$. Given that $\dot{\xi}_i(\tau)\in H_i(\xi(\tau))= \{ \nu_{i}^{M} \; (\psi_i (M) - \xi_i(\tau))\}$, due to the definition of $S(x)$, it follows that $\xi_i(\tau) = x_i < \psi_i (M)$. Similarly, $\varphi_i(\sigma) = y_i \geq \psi_i (M)$. But then, by Definition~\ref{df:flow_domain}, $x$ and $y$ are not located in the same flow domain $D$, contrary to the assertion of the theorem. Therefore, $1 \in S_i(x)$ implies $1 \in S_i(y)$. The converse is shown in the same way.

(b) $M \in \mathcal{M}_s$ and $\Psi(M) = \{ \}$. No solution remains in $D$, so by definition $S(x) =\linebreak S(y) = \{ \}$.

(c) $M \in \mathcal{M}_s$ and $\Psi(M) \not = \{ \}$. For every $i \in \{ 1, \ldots , n\}$ such that $D_i$ is located in a threshold or focal hyperplane, we must have $\dot{\xi}_i(t) = 0$ for solutions remaining in $D$. Consequently, $S_i(x)=\{0\}$ (idem $S_i(y)=\{0\}$). For all other $i$, suppose $1 \in S_i(x)$, but $1 \not \in S_i(y)$ (the argument for $0$ and $-1$ is similar). It follows that $\dot{\xi}_i(\tau)>0$ for some $\xi(t) \in \Xi_\Sigma$ in $D$ and $\tau \geq 0$, such that $\xi(\tau)=x$. However, $\dot{\varphi}_i(\sigma) \leq 0$ for all $\varphi(t) \in \Xi_\Sigma$ in $D$ and $\sigma \geq 0$, such that $\varphi(\sigma)=y$. From $\dot{\xi}_i(\tau)\in H_i(\xi(\tau))= \rect(\{ \nu_{i}^{M'} \; (\psi_i (M') - \xi_i(\tau)) \mid M' \in R(M)\})$, we conclude that $\xi_i(\tau) = x_i < \psi_i (M')$, for some $M' \in R(M)$. Similarly, we must have $\dot{\varphi}_i(\sigma)\in H_i(\varphi(\sigma))= \rect(\{ \nu_{i}^{M'} \; (\psi_i (M') - \varphi_i(\sigma)) \mid M' \in R(M)\})$. Now, if there were some $M' \in R(M)$ such that $\varphi_i(\sigma) = y_i < \psi_i (M')$, then by Lemma~\ref{lem:monotone-singular-existence} there would exist a solution such that $\dot{\varphi}_i(\sigma) > 0$. Since this contradicts the assumption, we conclude that $\varphi_i(\sigma) = y_i \geq \psi_i (M')$, for all $M' \in R(M)$. This implies, again by Definition~\ref{df:flow_domain}, that $x$ and $y$ are not located in the same flow domain $D$. Therefore, $1 \in S_i(x)$ implies $1 \in S_i(y)$ (and conversely).
\end{proof}

Notice that the definition of $S$ as a set is a direct consequence of the use of differential inclusions. Since the solutions of differential inclusions are not unique, several solutions may pass through $x$ and their derivatives may have a different sign in some dimension $i\in \{ 1, \ldots ,n\}$. This situation does not occur in our two-gene example.

Theorem~\ref{tm:abstraction} suggests that the partition of the state space introduced in this section can be used as an abstraction criterion better-adapted to the available experimental data on gene expression. This idea will be further developed in the next section.

\section{Qualitative abstraction of the dynamics of PA systems}
\label{sec:abstraction}

\subsection{Qualitative PA transition systems}
\label{sec:abstraction,qualPA}

As a preparatory step, we define a \textit{continuous transition system} having the same reachability properties as the original PA system $\Sigma$. Consider $x\in D$ and $x'\in D'$, where $D, D' \in \mathcal{D}$ are flow domains. If there exists a solution $\xi(t)$ of $\Sigma$ passing through $x$ at time $\tau \in \mathbb{R}_{\geq 0}$ and reaching $x'$ at time $\tau' \in \mathbb{R}_{>0}\cup \{\infty\}$, without leaving $D\cup D'$ in the time interval $[\tau, \tau']$, then the absolute continuity of $\xi(t)$ implies that $D$ and $D'$ are either equal or contiguous. More precisely, one of the following three cases holds: $D=D'$,  $D\subseteq \partial D'$, or $D'\subseteq \partial D$. We consequently distinguish three corresponding types of continuous transitions: \textit{internal}, denoted by $x\stackrel{int}{\longrightarrow} x'$, \textit{dimension increasing}, denoted by $x\stackrel{dim^+}{\longrightarrow} x'$, and \textit{dimension decreasing}, denoted by
$x\stackrel{dim^-}{\longrightarrow} x'$. The latter two terms refer to the increase or decrease in dimension of the flow domain when going from $D$ to $D'$. This leads to the following definition:

\begin{df}[PA transition system]\rm
\label{df:concreteTS}
$\Sigma\mbox{-}\mathrm{TS}=(\Omega, L, \Pi, \rightarrow, \models)$ is the transition system corresponding to the PA system $\Sigma=(\Omega, \Theta, H)$,  where:

\begin{enumerate}
\item $\Omega$ is the state space;
\item $L=\{int, dim^+, dim^-\}$ is a set of labels denoting the three different types of transitions;
\item $\Pi=  \{\Dsign= S \mid S \in 2^{\{-1, 0,1\}^n}\}$ is a set of propositions describing the derivative sign pattern of the concentration variables;
\item $\rightarrow$ is the transition relation describing the continuous evolution of the system, defined by $\rightarrow \subseteq \Omega \times L \times \Omega$, such that $x\stackrel{l}{\rightarrow} x'$ if and only if there exists $\xi(t) \in \Xi_\Sigma$ and $\tau, \tau'$, such that $0 \leq \tau<\tau'$, $\xi (\tau)=x$, $\xi(\tau')=x'$, and
  \begin{itemize}
   \item if $l= int$, then for all $t \in [\tau,\tau']$: $\xi(t)\in \flow(x)=\flow(x')$,
  \item if $l= dim^+$, then for all $t \in (\tau,\tau']$: $\xi(t)\in \flow(x') \neq \flow(x)$,
  \item if $l= dim^-$, then for all $t \in [\tau,\tau')$: $\xi(t)\in \flow(x) \neq \flow(x')$;
  \end{itemize}
\item $\models$ is the satisfaction relation of the propositions in $\Pi$, defined by  $\models \subseteq \Omega \times \Pi$, such that $x \models \Dsign= S$ if and only if $S= S(x)$.
\end{enumerate}
\end{df}

The satisfaction relation $\models$ thus associates to each point $x$ in the state space a qualitative description of the dynamics of the system at $x$. We define any sequence of points in $\Omega$, $(x^0, \ldots, x^m)$, $m \geq 0$, as a
\textit{run} of $\Sigma\mbox{-}\mathrm{TS}$ if for all $j \in \{ 0, \ldots , m-1\}$, there exists some $l\in L$ such that $x^{j}\stackrel{l}{\rightarrow} x^{j+1}$.  It is not difficult to show that a PA system $\Sigma$ and its corresponding PA transition system $\Sigma\mbox{-}\mathrm{TS}$ have equivalent reachability properties (see the proof in Appendix~\ref{app:proofs}).

\begin{pr}[Reachability equivalence]\rm
For all $x, x' \in \Omega$, there exists a solution $\xi(t)$ of $\Sigma$ and $\tau, \tau'$, such that $0 \leq \tau \leq \tau'$, $\xi(\tau)=x$, and $\xi(\tau')=x'$ if and only if there exists a run $(x^0, \ldots, x^m)$ of  $\Sigma\mbox{-}\mathrm{TS}$ such that $x^0=x$ and  $x^m=x'$.
\end{pr}

The continuous PA transition system has an infinite number of states and transitions, which makes it difficult to verify dynamical properties of interest, for instance by means of conventional tools for model checking \cite{HdJ2003}. However, we can define a discrete transition system, with a finite number of states and transitions, which preserves important properties of the qualitative dynamics of the system. In order to achieve this, we introduce the equivalence relation ${\sim_{\Omega}} \, \subseteq \Omega \times \Omega$ induced by the partition $\mathcal{D}$ of the state space: $x {\sim_{\Omega}} x'$ if and only if $\flow(x)=\flow(x')$. From Theorem~\ref{tm:abstraction} it follows that ${\sim_{\Omega}}$ is \textit{proposition-preserving} \cite{HdJ2205,HdJ2370}, in the sense that for all $x,x' \in D$ and for all $\pi\in \Pi$, $x\models \pi$ if and only if $x'\models \pi$.

The discrete or \textit{qualitative abstraction} of a PA transition system $\Sigma\mbox{-}\mathrm{TS}$, called \textit{qualitative PA transition system}, is now defined as the quotient transition system of $\Sigma\mbox{-}\mathrm{TS}$, given the equivalence relation ${\sim_{\Omega}}$ \cite{HdJ2205,HdJ2370}.

\begin{df}[Qualitative PA transition system]\rm
\label{df:abstractTS}
The qualitative PA transition system \linebreak corresponding to the PA transition system $\Sigma\mbox{-}\mathrm{TS}= (\Omega, L, \Pi, \rightarrow, \models)$ is $\Sigma\mbox{-}\mathrm{QTS}= \linebreak(\Omega/_{\sim_{\Omega}}, L, \Pi, \rightarrow_{\sim_{\Omega}}, \models_{\sim_{\Omega}})$.
\end{df}

\begin{pr}[Qualitative PA transition system]\rm
\label{prop:abstractTS}
Let $\Sigma\mbox{-}\mathrm{QTS} = (\Omega/_{\sim_{\Omega}}, L,$ $\Pi, \rightarrow_{\sim_{\Omega}}, \models_{\sim_{\Omega}})$ be the qualitative PA transition system corresponding to the PA transition system $\Sigma\mbox{-}\mathrm{TS}=(\Omega, L, \Pi, \rightarrow, \models)$. Then
\begin{enumerate}
\item $\Omega/_{\sim_{\Omega}} = \mathcal{D}$;
\item $\rightarrow_{\sim_{\Omega}} \, \subseteq \mathcal{D} \times L \times \mathcal{D}$, such that  $D\stackrel{l}{\rightarrow}_{\sim_{\Omega}} D'$ if and only if there exists $\xi(t) \in \Xi_\Sigma$ and $\tau, \tau',\; 0 \leq \tau<\tau'$, such that $\xi(\tau)\in D$, $\xi(\tau')\in D'$, and
  \begin{itemize}
  \item if $l= int$, then for all $t \in [\tau,\tau']$: $\xi(t)\in D=D'$,
  \item if $l= dim^+$, then for all $t \in (\tau,\tau']$: $\xi(t)\in D' \neq D$,
  \item if $l= dim^-$, then for all $ t \in [\tau,\tau')$: $\xi(t)\in D \neq D'$;
  \end{itemize}
\item $\models_{\sim_{\Omega}} \subseteq \mathcal{D} \times \Pi$, such that $D \models \Dsign= S$ if and only if for all $x\in D$, $S(x)=S$.
\end{enumerate}
\end{pr}

\begin{proof}
First, the quotient space $\Omega/_{\sim_\Omega}$ equals $\mathcal{D}$ by the definition of the equivalence relation $\sim_\Omega$. The second part of the proposition follows from the definition of $\rightarrow_{\sim_{\Omega}}$ as a relation on $\Omega/_{\sim_\Omega} \times L \times \Omega/_{\sim_\Omega}$ such that  $D\stackrel{l}{\rightarrow_{\sim_{\Omega}}} D'$ if and only if there exist $x \in D$ and $x' \in D'$ such that $x \stackrel{l}{\rightarrow} x'$ \cite{HdJ2205}. The expression for $\rightarrow_{\sim_{\Omega}}$ is a direct consequence of $\Omega/_{\sim_\Omega}=\mathcal{D}$ and Definition~\ref{df:concreteTS}. Third, $\models_{\sim_\Omega}$ is a relation defined on $\Omega/_{\sim_\Omega} \times \Pi$ such that $D \models_{\sim_\Omega} \pi$ if and only if there exists $x \in D$ such that $x \models \pi$ \cite{HdJ2205}. The properties considered here are of the type $\Dsign= S$ (Definition~\ref{df:concreteTS}). Theorem~\ref{tm:abstraction} implies the invariance of $S$ for all $x \in D$.
\end{proof}

Notice that the transitions labeled by $dim^+$ or $dim^-$ connect two different flow domains, since in Proposition~\ref{prop:abstractTS} we require that $D \neq D'$. This corresponds to a continuous evolution of the system along which it switches from one flow domain to another. On the contrary, the transitions labeled by $int$ correspond to the continuous evolution of the system in a single flow domain.

As for $\Sigma\mbox{-}\mathrm{TS}$, we define any sequence of flow domains $(D^0, \ldots, D^m)$, $m \geq 0$, as a \textit{run} of $\Sigma\mbox{-}\mathrm{QTS}$ if and only if for all $j \in \{0, \ldots ,m-1\}$, there exists $l\in L$ such that \linebreak $D^{j}\stackrel{l}{\rightarrow}_{\sim_{\Omega}} D^{j+1}$.  The satisfaction relation $\models_{\sim_{\Omega}}$ associates to every run a qualitative description of the evolution of the derivatives over time. $\Sigma\mbox{-}\mathrm{QTS}$ can be represented by a directed graph $G=(\mathcal{D}, \rightarrow_{\sim_{\Omega}})$, called the \textit{state transition graph}. The paths in $G$ represent the runs of the system. The state transition graph corresponding to the two-gene example is represented in Figure~\ref{fig:partition_singular2}(b), and $(D^{1.1}, D^{2.2}, D^{3.2}, D^{4.2}, D^{4.1})$ is an example of a run.

It directly follows from the definitions of quotient transition system and simulation of transition systems \cite{HdJ2205,HdJ2370} that $\Sigma\mbox{-}\mathrm{QTS}$ is a \textit{simulation} of $\Sigma\mbox{-}\mathrm{TS}$. The converse is not true in general, because $\Sigma\mbox{-}\mathrm{QTS}$ and $\Sigma\mbox{-}\mathrm{TS}$ are not bisimilar.

\begin{pr}\rm
\label{lem:simul}
$\Sigma\mbox{-}\mathrm{QTS}$ is a simulation of $\Sigma\mbox{-}\mathrm{TS}$.
\end{pr}

As a consequence of Proposition~\ref{lem:simul}, if there exists a run $(x^0, \ldots, x^m)$ of  $\Sigma\mbox{-}\mathrm{TS}$, then there also exists a run $(D^0, \ldots, D^{m})$ of $\Sigma\mbox{-}\mathrm{QTS}$ such that $x^i \in D^i$, for all $i \in \{0, \ldots , m\}$. In other words, $\Sigma\mbox{-}\mathrm{QTS}$ is a \textit{conservative approximation} of $\Sigma\mbox{-}\mathrm{TS}$.

\subsection{Invariance of qualitative PA transition systems in parameter space}

$\Sigma\mbox{-}\mathrm{QTS}$ provides a qualitative picture of the dynamics of a genetic regulatory network. This picture generally depends on the values of the parameters in the PADE model. Since exact numerical values for the thresholds $\theta$ and the production and degradation constants $\kappa$ and $\gamma$ are usually not available, it is important to verify to which extent $\Sigma\mbox{-}\mathrm{QTS}$ is robust to variations in the parameter values.

We therefore introduce a second equivalence relation $\sim_{\Gamma} \, \subseteq \Gamma \times \Gamma$, defined on the \textit{parameter space} $\Gamma \subseteq \mathbb{R}_{> 0}^{q}$ of the PA system, with $q$ the number of parameters in $\Sigma$. Two parameter vectors $p,p' \in \Gamma$ are equivalent, if their corresponding qualitative PA transition systems, and hence their state transition graphs, are isomorphic. Given the equivalence relation $\sim_\Gamma$, we denote by $\Gamma/_{\sim_\Gamma}$ the quotient parameter space. That is, $\Gamma/_{\sim_\Gamma}$ is a partition of the parameter space consisting of sets over which the qualitative PA transition system is invariant.

For our purpose, subsets of $\Gamma$ defined by the following \textit{parameter inequality constraints} are particularly interesting.

\begin{df}[Parameter inequality constraints]\rm
\label{df:paraminequal}
The parameter inequality constraints of $\Sigma$ are a set of total strict orders on $\{\theta_i^1, \ldots, \theta_i^{p_i}\} \cup \{\psi_i(M) \mid M\in \mathcal{M}_r\}$, for every $i\in \{ 1, \ldots ,n\}$.
\end{df}

The following theorem states that $\Sigma\mbox{-}\mathrm{QTS}$ is invariant over the subsets of $\Gamma$ defined by the inequality constraints of Definition~\ref{df:paraminequal}. That is, the qualitative dynamics of $\Sigma$ is robust to changes in parameter values that do not change the total order of the thresholds and focal concentrations.

\begin{tm}[Invariance]\rm
\label{tm:mod_equiv}
Let $P\subseteq \Gamma$ be a set defined by the parameter inequality constraints of $\Sigma$. Then, there exists some $Q \in \Gamma/_{\sim_\Gamma}$ such that $P\subseteq Q$.
\end{tm}

\begin{proof}
Let $p,p' \in P$ be two parameter vectors. Given that $p$ and $p'$ satisfy the same parameter inequality constraints, they lead to the same mode and flow domain partitions (Definitions~\ref{df:mode_domain} and \ref{df:flow_domain}), and thus the same quotient space $\Omega/_{\sim_{\Omega}} = \mathcal{D}$. As will be seen in the next section, the relations $\rightarrow_{\sim_{\Omega}}$ and $\models_{\sim_{\Omega}}$ can be characterized by means of rules involving inequality constraints (Propositions~\ref{tm:Dsign_a} to \ref{tm:qual_tr_b}). It can be directly verified that the necessary and sufficient conditions in the rules apply in exactly the same way to $p$ and $p'$.
\end{proof}

Suppose that protein A inhibits the expression of gene \textit{b} at a concentration lower than that required for the inhibition of the expression of its own gene. This gives the inequality constraint $\theta_{a}^{1} < \theta_{a}^{2}$ in Figure~\ref{fig:partition_singular2}(c). Moreover, if we further assume that when gene \textit{a} is active, the concentration of protein A tends towards a level above which autoinhibition occurs, we obtain the inequality constraint $\kappa_{a}/\gamma_{a} > \theta_{a}^{2}$. For these inequality constraints, and similar constraints for protein B, the state transition graph in Figure~\ref{fig:partition_singular2}(b) is invariant. Whereas exact numerical values for the parameters are usually not available, the weaker information required for the formulation of the inequality constraints can often be obtained from the experimental literature. This will be illustrated in Section~\ref{sec:application} for the \textit{E. coli} example.

\section{Symbolic computation of qualitative PA transition system}
\label{sec:computation}

The computation of the qualitative PA transition system $\Sigma\mbox{-}\mathrm{QTS}$ is greatly simplified by the fact that the domains $D$ and the focal sets $\Psi(M)$ are hyperrectangular, which allows them to be expressed as product sets, {\it i.e.} $D = D_1 \times \ldots \times D_n$ and $\Psi(M) = \Psi_1(M)\times \ldots \times \Psi_n(M)$. As a consequence, the computation can be carried out for each dimension separately. In this section, we will describe rules to determine the set of states $\mathcal{D}$, the satisfaction relation $\models_{\sim_\Omega}$, and the transition relation $\stackrel{}{\rightarrow}_{\sim_{\Omega}}$, as well as their implementation in the computer tool GNA. In practice the computations reduce to simple checks of ordering relations, which can be carried out symbolically by means of the parameter inequality constraints.

\subsection{Computation of states}
\label{sec:computation,domains}

In order to compute the states of $\Sigma\mbox{-}\mathrm{QTS}$, we need to determine the flow domain partition $\mathcal{D}$ of $\Omega$ (Proposition~\ref{prop:abstractTS}). This requires a total ordering of the threshold concentrations $\{\theta_i^1, \ldots, \theta_i^{p_i}\}$ and the focal concentrations $\{\psi_i(M) \mid M\in \mathcal{M}_r\}$, $i\in \{ 1, \ldots ,n\}$ (Definition~\ref{df:flow_domain}). The parameter inequality constraints of $\Sigma$ provide this information.

Each flow domain has associated to it a number of properties, defined by the satisfaction relation $\models_{\sim_\Omega}$. From Proposition~\ref{prop:abstractTS} it follows that computing the relation $\models_{\sim_\Omega}$ amounts to computing $S(x)$. The following two propositions, which are direct consequences of the basic mathematical properties of solutions of PA systems discussed in Section~\ref{sec:mathanal}, show how to compute the derivative sign patterns.

\begin{pr}[Computation of $\Dsign$]\rm
\label{tm:Dsign_a}
Let $D\in \mathcal{D}$ be an instantaneous flow domain. $D \models_{\sim_\Omega} \Dsign=\{\}$.
\end{pr}

\begin{proof}
There exists no solution remaining in $D$ for some time, so by the definition of $S(x)$, we have $S(x) = \{\}$, for every $x \in D$.
\end{proof}

\begin{pr}[Computation of $\Dsign$]\rm
\label{tm:Dsign_b}
Let $D\in \mathcal{D}$ be a persistent flow domain. \linebreak $D \models_{\sim_\Omega} \Dsign = S \not = \{ \}$, $S = S_1 \times \ldots \times S_n$, and for every $i \in \{ 1, \ldots ,n\}$,
\begin{description}
\item[] if $D_i$ is located in a threshold or focal concentration hyperplane, then $S_i=\{0\}$;
\item[] otherwise,
\begin{itemize}
\item $-1 \in S_i, \mbox{ if and only if for all } x\in D \mbox{ there exists } \psi\in \Psi(\mode(D)) \mbox{ such that } \linebreak \psi_i- x_i <0$;
\item $0 \in S_i, \mbox{ if and only if for all } x\in D \mbox{ there exists } \psi \in \Psi(\mode(D)) \mbox{ such that } \linebreak \psi_i- x_i =0$;
\item $1 \in S_i, \mbox{ if and only if for all } x\in D \mbox{ there exists } \psi \in \Psi(\mode(D)) \mbox{ such that } \linebreak \psi_i- x_i >0$.
\end{itemize}
\end{description}
\end{pr}

\begin{proof}
Let $M = \mode(D)$. Because $D$ is persistent, there exists a solution $\xi(t) \in \Xi_\Sigma$ remaining in $D$ for some time. Let $\xi(\tau) = x \in D$ for some $\tau \geq 0$. From (\ref{eq:S}) we infer $S(x) \not = \{ \}$. By Theorem~\ref{tm:abstraction} this must hold for all $x \in D$, so $S \not = \{ \}$.

Let $D_i$ be located in a threshold or focal concentration hyperplane. For every solution $\xi(t)$ in $D$, such that $\xi(\tau) = x \in D$ for some $\tau \geq 0$, the derivative of $\xi(\tau)$ exists, and $\dot{\xi}(\tau) \in H(\xi(\tau))$, it holds that $\dot{\xi}_i(\tau)=0$ and hence $S_i(x) = \{0\}$. By Theorem~\ref{tm:abstraction} this must hold for all $x \in D$, so $S_i = \{0\}$.

Alternatively, $D_i$ is not located in a threshold or focal concentration hyperplane. We only consider the case $-1 \in S_i$ (the argument for 0 and 1 is similar). We first prove necessity of the inequality condition. $-1 \in S_i$ means that for all $x \in D$, there exists a solution $\xi(t) \in \Xi_\Sigma$ with $\xi(\tau) = x$ and $\tau \geq 0$, such that $\dot{\xi}_i(\tau)<0$ (Proposition~\ref{prop:abstractTS} and (\ref{eq:S})). If $M$ is regular, then $H_i(\xi(\tau))= \{ \nu_{i}^{M} \; (\psi_i (M) - \xi_i(\tau))\}$. Since $\dot{\xi}_i(\tau)\in H_i(\xi(\tau))$, it follows that $\psi_i(M) - x_i < 0$. If M is singular, then $H_i(\xi(\tau))=\rect(\{ \nu_{i}^{M'} \; (\psi_i (M') - \xi_i(\tau)) \mid M' \in R(M)\})$. $\dot{\xi}_i(\tau)<0$ implies that there exists some $\psi(M')$, $M' \in R(M)$, such that $\psi_i (M') - x_i < 0$. Conversely, by Lemmas~\ref{lem:monotone-regular} and \ref{lem:monotone-singular-existence} there exist a solution $\xi(t) \in \Xi_\Sigma$ in $M$ monotonically converging from $x$ to $\psi$, with $\xi(0) = x$. Because $\psi_i - x_i < 0$, we have $\dot{\xi}_i(t) < 0$, $t \geq 0$, and thus $-1 \in S_i(x)$. By Theorem~\ref{tm:abstraction} it follows that $-1 \in S_i$, which proves sufficiency.
\end{proof}

Notice that the total ordering on the threshold and focal concentrations expressed by the parameter inequality constraints (Definition~\ref{df:paraminequal}) allows one to decide which of the conditions in the second part of Proposition~\ref{tm:Dsign_b} are satisfied. As a prerequisite, $\Psi(\mode(D))$ needs to be determined, which is also straightforward, even for singular mode domains, given the definition of the focal set and the parameter inequality constraints. We illustrate the application of Propositions~\ref{tm:Dsign_a} and \ref{tm:Dsign_b} to our two-gene example.

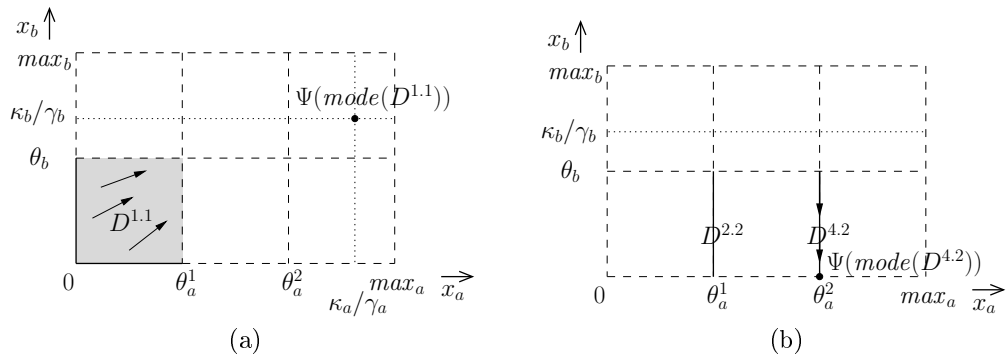
\begin{figure*}[htb]
  \begin{minipage}[b]{14.2cm}
    \centering
    \begin{minipage}[b]{7cm}
      \begin{center}
        \scalebox{.55}{\input{figures/sign_M1.pstex_t}}\\
        \mbox{(a)}
      \end{center}
    \end{minipage}
    \begin{minipage}[b]{7cm}
      \begin{center}
        \scalebox{.55}{\input{figures/sign_M11.pstex_t}}\\
        \mbox{(b)}
      \end{center}
    \end{minipage}
  \end{minipage}
  \caption{\label{fig:sign} Derivative signs of flow domains: (a)  $\Dsign(D^{1.1})=\{(1, 1)'\}$ and (b) $\Dsign(D^{4.2})=\{(0, -1)'\}$ and $\Dsign(D^{2.2})=\{\}$.}
\end{figure*}

First, consider the flow domain $D^{1.1}$ in Figure~\ref{fig:sign}(a). $\Psi(\mode(D^{1.1}))$ equals \linebreak $\{(\kappa_a/\gamma_a, \kappa_b/\gamma_b)'\}$, so that $D^{1.1}$ is persistent. We therefore apply Proposition~\ref{tm:Dsign_b} to determine $S = S_a \times S_b$, such that $D^{1.1} \models_{\sim_\Omega} \Dsign=S$. Given that $D_a^{1.1}=[0, \theta_a^1)$ and $\psi_a(\mode(D^{1.1}))=\kappa_a/\gamma_a$, and $\theta_a^1 < \kappa_a/\gamma_a$ according to the parameter inequality constraints in Figure~\ref{fig:partition_singular2}(c), it follows that $\psi_a-x_a>0$, for all $x \in D^{1.1}$ and $\psi \in \Psi(\mode(D^{1.1}))$. Consequently, we infer $S_a=\{1\}$. Repeating this procedure in the $x_b$-dimension, we similarly find $S_b=\{1\}$, so that $D^{1.1} \models_{\sim_\Omega} \Dsign=\{(1,1)'\}$. This means that the solutions in $D^{1.1}$ are strictly increasing in both dimensions.

As a second example, consider the flow domain $D^{4.2}$, represented in Figure~\ref{fig:sign}(b). $M^4=\mode(D^{4.2})$ is a singular mode domain, so we first have to compute $\Psi(M^4)$. It can be immediately verified that $R(M^4)=\{ M^3, M^5\}$, where $\psi(M^3)=(\kappa_a/\gamma_a, 0)'$ and $\psi(M^5)=(0,0)'$ (Figure~\ref{fig:exnet-coarse_phase_space}(b)). As a consequence, $\rect(\{\psi(M^3), \psi(M^5)\}) = [0, \kappa_a/\gamma_a] \times \{0\}$. From the parameter inequality constraints in Figure~\ref{fig:partition_singular2}(c) it follows that $0<\theta_a^2< \kappa_a/\gamma_a$, so that $\supp(M^{4})$ and $\rect(\{\psi(M^3), \psi(M^5)\})$ intersect at $\Psi(M^4) = \{ (\theta_a^2,0)'\}$. $D^{4.2}$ is persistent, so that Proposition~\ref{tm:Dsign_b} applies. We determine $S = S_a \times S_b$, such that $D^{4.2} \models_{\sim_\Omega} \Dsign=S$. Since $D_a^{4.2}$ coincides with a threshold hyperplane, $S_a=\{0\}$. Bearing in mind that $D_b^{4.2}=(0, \theta_b)$ and $\psi_b(M^{4})=0$, we find $S_b=\{-1\}$. This results in  $D^{4.2} \models_{\sim_\Omega} \Dsign=\{(0,-1)'\}$, which is of course consistent with the fact that the solutions sliding along $D^{4.2}$ monotonicaly converge towards $\Psi(M^4)$, and are therefore strictly decreasing in the $x_b$-dimension.

\subsection{Computation of transitions between states}
\label{sec:computation,transitions}

In Section~\ref{sec:abstraction,qualPA} we have distinguished three types of transitions: $int$, $dim^-$, and $dim^+$. We will formulate, for each of these three cases, transition rules that can be applied by means of the parameter inequality constraints of $\Sigma$.

The transitions of type $int$ are easy to determine, since by Proposition~\ref{prop:abstractTS} they are necessarily self-transitions, from a flow domain $D$ to itself, which occur if and only if $D$ is persistent.

\begin{pr}[Computation of $int$ transition]\rm
\label{tm:qual_tr_z}
Let $D \in \mathcal{D}$. $D \stackrel{int}{\longrightarrow}_{\sim_\Omega} D$ if and only if $D$ is persistent.
\end{pr}

\begin{proof}
By Proposition~\ref{prop:abstractTS} an $int$ transition occurs if and only if there exist solutions $\xi(t) \in \Xi_\Sigma$ remaining in $D$. That is, if and only if $D$ is persistent.
\end{proof}

Recall that the persistence of a domain can be determined by checking whether \linebreak $\Psi(\mode(D)) \not = \{ \}$ (Lemmas~\ref{lem:monotone-regular} and \ref{lem:instantaneous-singular}). In our two-gene example, the focal set $\Psi(M^{1})$ is not empty. Therefore, $D^{1.1}$ is persistent and there exists an $int$ transition on $D^{1.1}$. On the other hand, $\Psi(M^{2})$ is empty, so $D^{2.2}$ is instantaneous and does not have an $int$ transition (Figure~\ref{fig:partition_singular2}(b)).

A $dim^+$ transition $D \stackrel{dim^+}{\longrightarrow}_{\sim_\Omega} D'$ is dimension increasing, and therefore requires that $D\subseteq \partial D'$ (Section~\ref{sec:abstraction,qualPA}). In order to make $D'$ reachable from $D$, the solutions in $D'$ must point away from $D$ in the dimensions $i \in \{1, \ldots ,n\}$ for which $D_i \subseteq \partial D'_i$. This is expressed by the following proposition.

\begin{pr}[Computation of $dim^+$ transition] \rm
\label{tm:qual_tr_a}
Let $D,D' \in \mathcal{D}$ and $D \subseteq \partial D'$. \linebreak $D \stackrel{dim^+}{\longrightarrow}_{\sim_\Omega} D'$ if and only if $\Psi(\mode(D')) \not = \{\}$ and there exist $x\in D$, $x'\in D'$, and $\psi'\in \Psi(\mode(D'))$, such that for all $i\in \{1, \ldots ,n\}$ for which $D_i \subseteq \partial D'_i$, it holds that
\begin{equation}
(\psi'_i- x_i)(x_i'-x_i)>0.
\end{equation}
\end{pr}

\begin{proof}
Let $M' = \mode(D')$. We first prove necessity by contraposition. If $\Psi(M') = \{\}$, then no solutions remain in $D'$ and a transition from $D$ to $D'$ is not possible. Otherwise, suppose that for all $x\in D$, $x'\in D'$, and $\psi'\in \Psi(M')$, there exists some $i\in \{ 1, \ldots ,n\}$ for which $D_i \subseteq \partial D'_i$ and $(\psi'_i- x_i)(x_i'-x_i) \leq 0$. Furthermore, assume $x_i'-x_i>0$ for all $x\in D$, $x'\in D'$ (the case $x_i'-x_i<0$ goes analogously). As a consequence, $\psi'_i \leq x_i < x'_i$, for all $x\in D$, $x' \in D'$, and $\psi'\in \Psi(M')$. By Lemmas~\ref{lem:monotone-regular} and \ref{lem:monotone-singular}, for all solutions $\xi (t) \in \Xi_\Sigma$ in $D'$, $\xi_i (t)$ monotonically converges towards $\Psi_i(M')$. As a consequence, no solution enters $D'$ from $D$ and there does not exist a transition $D \stackrel{dim^+}{\longrightarrow}_{\sim_\Omega} D'$.

Next, we prove sufficiency. Suppose $\Psi(M') \not = \{\}$ and there exist $x\in D$, $x'\in D'$, and \linebreak $\psi'\in \Psi(M')$, such that for all $i\in \{1, \ldots ,n\}$ for which $D_i \subseteq \partial D'_i$, it holds that $(\psi'_i- x_i)(x_i'-x_i)>0$. By the definition of the flow domain partition (Definition~\ref{df:flow_domain}) this holds for all $x\in D$ and $x'\in D'$. We further assume that $x_i'-x_i>0$ for all $x\in D$, $x'\in D'$ (the case $x_i'-x_i<0$ goes analogously). As a consequence, $x_i < x'_i < \psi'_i$ for all $x\in D$ and $x'\in D'$, and some $\psi' \in \Psi(M')$. By Lemma~\ref{lem:monotone-singular-existence}, for all $x' \in D'$ there exists a solution $\xi (t) \in \Xi_\Sigma$ in $D'$, with $\xi (0) = x'$, which monotonically converges towards $\psi'$. More precisely, by the construction of $\xi(t)$ in the proof of Lemma~\ref{lem:monotone-singular-existence}, we have $\dot{\xi}_i(t) > 0$ as long as $\xi_i (t) < \phi'_i$. As a consequence, it is possible for solutions to enter $D'$ from $D$ and there exists a transition $D \stackrel{dim^+}{\longrightarrow}_{\sim_\Omega} D'$.
\end{proof}

The total strict ordering defined by the parameter inequality constraints allows the necessary and sufficient conditions for the existence of a $dim^+$ transition to be straightforwardly tested.

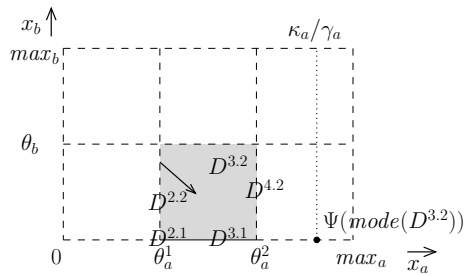
\begin{figure}[htb]
\centering
          \scalebox{.5}{\input{figures/comp_trans_M3.pstex_t}}
  \caption{\label{fig:trans_a} $dim^+$ transitions between flow domains. Representation of the state space, with  a trajectory entering $D^{2.2}$ from $D^{3.2}$.}
\end{figure}

As an illustration of the proposition, consider the $dim^+$ transition from $D^{2.2}$ to $D^{3.2}$ (Figure~\ref{fig:trans_a}). Notice that $D^{2.2} \in  \partial D^{3.2}$. Moreover, $D_a^{2.2} \in  \partial D_a^{3.2}$, so that $x'_a-x_a>0$, for all $x\in D^{2.2}$ and $x'\in D^{3.2}$. Since $\psi_a(\mode(D^{3.2}))=\kappa_a/\gamma_a$, and $\theta_a^2<\kappa_a/\gamma_a$ due to the parameter inequality constraints (Figure~\ref{fig:partition_singular2}(c)), we have $\psi'_a-x_a'>0$, for all $x'\in D^{3.2}$ and $\psi' \in \Psi(\mode(D^{3.2}))$. As a consequence, Proposition~\ref{tm:qual_tr_a} allows us to infer that $D^{2.2} \stackrel{dim^+}{\longrightarrow}_{\sim_\Omega} D^{3.2}$. In a similar way, we can infer that there is a $dim^+$ transition from $D^{2.1}$ to $D^{3.1}$. However, as expected from the direction of the flow in $\Sigma$, a $dim^+$ transition from $D^{2.1}$ to $D^{3.2}$ is excluded, since $\psi'_b-x'_b<0$ and $x'_b -x_b>0$, for all $x \in D^{2.1}$, $x' \in D^{3.2}$ and $\psi' \in \Psi(\mode(D^{3.2}))$.

The rule for $dim^-$ transitions is almost symmetric and given in Appendix~\ref{app:proofs}.

\subsection{Computer implementation}
\label{sec:computation,implementation}

In summary, given a PADE system $\Sigma$  and parameter inequality constraints, we can compute its qualitative abstraction, that is, the qualitative PA transition system $\Sigma\mbox{-}\mathrm{QTS}$, by means of Propositions~\ref{tm:Dsign_a} to \ref{tm:qual_tr_b}. Instead of numerically computing the derivative signs in the domains and the transitions between domains, we have developed symbolic algorithms exploiting the parameter inequality constraints of Definition~\ref{df:paraminequal}. In particular, we map the total strict order on $B_i = \{\theta_i^1, \ldots, \theta_i^{p_i}\} \cup \{\psi_i(M) \mid M\in \mathcal{M}_r\}$, $i \in \{ 1, \ldots ,n\}$, to the set $C_i = \{ 0, \ldots , |B_{i}| \nolinebreak +\nolinebreak1\}$. This allows $D_i$ and $\Psi_i(M)$ to be expressed as intervals on $C$. The conditions in the propositions then naturally translate into simple inequality tests on integer coordinates of domains and focal concentrations (see \cite{HdJ2538} for details).

The computation of $\Sigma\mbox{-}\mathrm{QTS}$ has been implemented in a new version of the computer tool {\it Genetic Network Analyzer (GNA)} \cite{HdJ2427}. In order to facilitate the analysis of $\Sigma\mbox{-}\mathrm{QTS}$, the state transition graph generated by GNA can be exported to standard model-checking tools like NuSMV and CADP \cite{HdJ2427}.

In practice, since the number of flow domains in the state space grows exponentially with the number of genes in the network, it is not usually possible to compute the complete state transition graph. However, knowledge of the part of the graph reachable from a (set of) initial flow domain(s) is often sufficient to answer most of the questions of biological interest. In GNA it is possible to either compute the complete qualitative PA transition system or carry out a \textit{reachability analysis} from a specified set of initial domains. Moreover, GNA identifies the equilibrium states, that is, the flow domains corresponding to (sets of) equilibrium points, by testing whether $D \models_{\sim_\Omega} 0\in \Dsign$. For each of the equilibrium states, the \textit{attractor set} is computed, that is, the set of states from which the equilibrium state is reachable.

\section{Qualitative analysis of nutritional stress response in \textit{E. coli}}
\label{sec:application}

In case of nutritional stress, an \textit{Escherichia coli} population abandons exponential growth and enters a non-growth state called {\it stationary phase}. This growth-phase transition is accompanied by numerous physiological changes in the bacteria, concerning among other things the morphology and the metabolism of the cell, as well as gene expression \cite{HdJ2336}. On the molecular level, the transition from exponential phase to stationary phase is controlled by a complex genetic regulatory network integrating various environmental signals. The molecular basis of the adaptation of the growth of \textit{E. coli} to nutritional stress conditions has been the focus of extensive studies for decades \cite{HdJ2338}. However, notwithstanding the large amount of information accumulated on the genes, proteins, and other molecules known to be involved in the stress adaptation process, there is currently no global understanding of how the response of the cell emerges from the network of molecular interactions. This suggest the use of modeling and simulation tools to study the dynamics of the stress response. However, with some exceptions \cite{HdJ2499}, numerical values for the parameters characterizing the interactions and the molecular concentrations are absent, which makes it difficult to apply traditional methods.

\begin{figure*}[htp]
\begin{minipage}[b]{14cm}
\begin{center}
\scalebox{0.50}{\input{figures/Ecoli_network.pstex_t}}
\end{center}
\end{minipage}
\caption{\label{fig:reseau} Network of key genes, proteins, and regulatory interactions involved in the nutritional stress network in \textit{Escherichia coli}. The contents of the boxes labelled `activation' and `supercoiling' are detailed in \protect\cite{HdJ2461}.}
\end{figure*}

The above circumstances have motivated the qualitative analysis of the nutritional stress response network in \textit{E. coli} by means of the method presented in this paper \cite{HdJ2461}. On the basis of literature data, we have decided to focus, as a first step, on a network of six genes that are believed to play a key role in the carbon starvation response (Figure~\ref{fig:reseau}). The network includes genes involved in the transduction of the nutritional stress signal (the global regulator {\it crp} and the adenylate cyclase {\it cya}), metabolism (the global regulator {\it fis}), cellular growth (the {\it rrn} genes coding for stable RNAs), and DNA supercoiling, an important modulator of gene expression (the topoisomerase {\it topA} and the gyrase {\it gyrAB}).

Based on this information, a PADE model of seven variables has been constructed, one protein concentration variable for each of the six genes and one input variable ($u_{\signal}$) representing the presence or absence of a carbon starvation signal \cite{HdJ2461}. As an illustration, the piecewise-affine differential equation and the parameter inequality constraints for the state variable $x_\topA$ are given below.
\begin{multline*}
\dot{x}_{\topA}=\; \kappa_{\topA}\; s^+(x_{\gyrAB}, \theta_{\gyrAB}^2)\; s^-(x_{\topA}, \theta_{\topA}^1)\; s^+(x_{\fis}, \theta_{\fis}^4) -\; \gamma_{\topA}\; x_{\topA}
\end{multline*}
\begin{equation*}
    0 < \theta_{\topA}^1 < \theta_{\topA}^2 < \kappa_{\topA} / \gamma_{\topA} < \maX_{\topA}
\end{equation*}

The equation and inequalities state that in the presence of a high concentration of Fis ($s^+(x_{\fis}, \theta_{\fis}^4)=1$), and of a high level of DNA supercoiling ($s^+(x_{\gyrAB}, \theta_{\gyrAB}^2) \linebreak \cdot s^-(x_{\topA}, \theta_{\topA}^1)=1$), the concentration of TopA increases, converging towards a high value ($\kappa_{\topA}/ \gamma_{\topA}> \theta_{\topA}^2$).

\begin{figure}[htbp]
  \centering
  \begin{center}
    \scalebox{0.5}{\input{figures/temp_profile_reduced.pstex_t}}
  \end{center}
  \caption{Temporal evolution of Fis and CRP concentrations in the run $(D^1, \ldots, D^{31})$. For every $D^j$ in the run, the profiles display the corresponding concentration intervals $D^j_{fis}$ and $D^j_{crp}$. The symbols $\uparrow$, $\downarrow$, and $\circ$ indicate the sign of the derivative for persistent states.}
  \label{fig:temp_profile}
\end{figure}
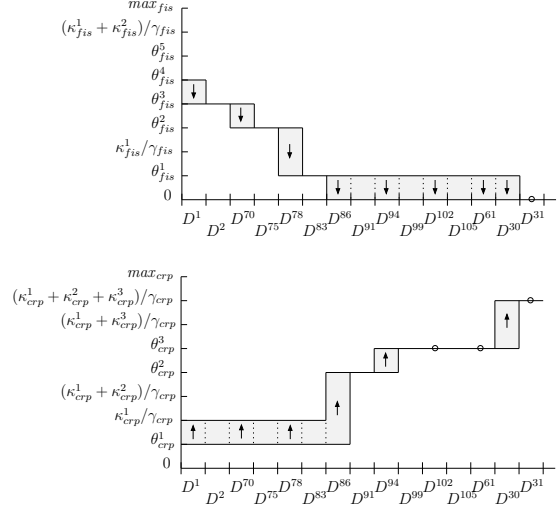

Using the computer tool GNA, we have performed reachability analyses on the qualitative PA transition system associated with the PADE model. The simulation of the entry into stationary phase has given rise to a state transition graph of 66 states, computed in less than 1 s on a PC (800\,MHz, 256\,Mb). Figure~\ref{fig:temp_profile} represents the temporal evolution of two of the protein concentrations in a run. The predicted expression profiles are consistent with the observations \cite{HdJ2461}.

The coupling of GNA with model-checking tools \cite{HdJ2427} has allowed a more systematic verification of observed dynamical properties. In \cite{HdJ2533}, the measured concentration of the global regulator Fis is shown to decrease and become steady in stationary phase, which is characterized by a low concentration $x_\rrn$ of stable RNAs. This can be expressed by means of the following CTL formula:
\begin{equation}
\op{EF} (\dot{x}_\fis <0 \, \wedge \op{EF} (\dot{x}_\fis = 0 \wedge x_\rrn < \theta_\rrn )).
\label{eq:CTL-fis}
\end{equation}
The verification of (\ref{eq:CTL-fis}) takes a fraction of a second to complete and shows that the formula holds for the state transition graph. The observation that the transcription of $\cya$ is negatively regulated by cAMP and CRP \cite{HdJ2534} is also reproduced by the model. In fact, the following CTL formula is satisfied by the graph:
\begin{multline}
\op{AG} (x_\crp>\theta_\crp^3 \wedge x_\cya>\theta_\cya^3 \wedge u_\signal > \theta_\signal \rightarrow \op{EF}\; \dot{x}_\cya <0).
\label{eq:CTL-cya}
\end{multline}
The formula expresses that always, for high levels of CRP and Cya, in the presence of the carbon starvation signal, the system can eventually reach a state in which the expression of $\cya$ decreases.

The application of the method has led to new insights into how the carbon starvation signal results in the slowing-down of bacterial growth characteristic for the stationary phase \cite{HdJ2461}. In summary, the analysis has brought to the fore the role of the mutual inhibition of Fis and CRP, which in the presence of a carbon starvation signal results in the inhibition of \textit{fis} and in the activation of \textit{crp}. This causes a decrease of the expression of the \textit{rrn} genes, which code for stable RNAs and are a reliable indicator of cellular growth. In addition to this increased understanding of the transition from exponential to stationary phase, the model has yielded predictions on the occurrence of damped oscillations in some of the protein concentrations after a nutrient upshift, predictions that are being tested in our laboratory.

We are currently working on extended models of the nutritional stress response network. The increase in the number of variables naturally leads to the generation of larger state transition graphs by our method. In order to investigate the upscalability of the method more systematically, we have applied it to a PADE model with nine state and two input variables, describing the initiation of sporulation in the bacterium \textit{Bacillus subtilis} \cite{HdJ2126}. More specifically, we have analyzed the response of the cell to carbon starvation in the case of the wild-type and a dozen of mutant strains. On average, the state transition graphs generated by our method consist of 585 states, with a maximum of 2234 states (computed in 14 s). The analysis of graphs of this size does not pose any problems for current model-checking tools, which shows that our approach is upscalable to large and complex networks.

\section{Discussion and conclusions}
\label{sec:discussion}

We have presented a method for the qualitative analysis of hybrid models of genetic regulatory networks. The method is based on a class of piecewise-affine differential equation models that has been well-studied in mathematical biology. By defining a qualitative abstraction preserving the sign pattern of the derivatives of concentration variables, the continuous PA transition system associated with a PADE model is transformed into a discrete or qualitative PA transition system whose properties can be analyzed by means of classical model-checking tools. The qualitative PA transition system is a conservative approximation of the underlying continuous PA transition system and can be easily computed in a symbolic manner by exploiting inequality constraints on the parameters. We have applied the implementation of the method to the analysis of a system whose functioning is not well-understood by biologists today, the nutritional stress response in the bacterium \textit{E. coli}.

\begin{figure*}[htp]
  \begin{minipage}[t]{14cm}
    \begin{minipage}[b]{4cm}
      \begin{center}
        \scalebox{0.45}{\input{figures/Phase_spaceEcoli_flow.pstex_t}}\\
        \hspace*{.85cm}\mbox{(a)}
      \end{center}
    \end{minipage}\hfill
    \begin{minipage}[b]{3.25cm}
      \begin{center}
        \scalebox{0.45}{\input{figures/Phase_spaceEcoliCoarse.pstex_t}}\\
        \hspace*{.75cm}\mbox{(b)}
      \end{center}
    \end{minipage}\hfill
    \begin{minipage}[b]{4.5cm}
      \begin{center}
        \scalebox{0.45}{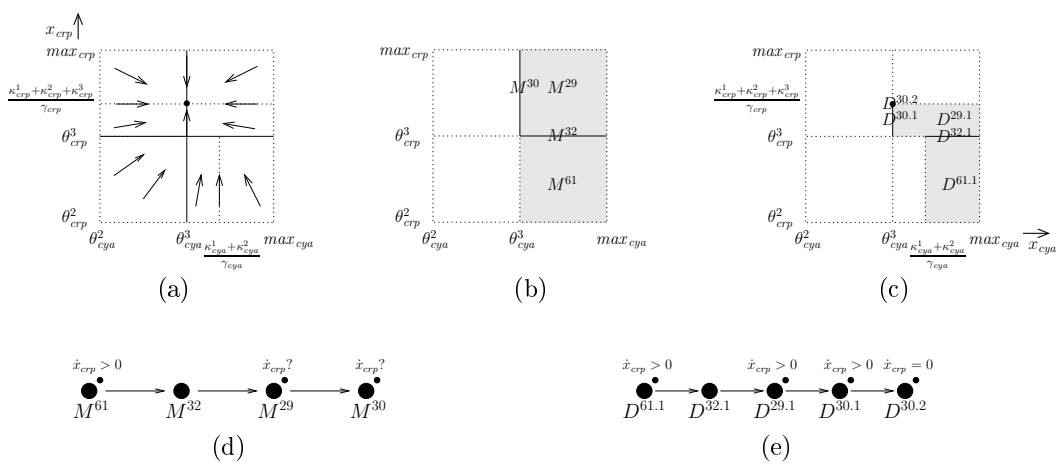}
        \hspace*{.3cm}\mbox{(c)}
      \end{center}
    \end{minipage}
  \end{minipage}\\
  \vspace*{1em}\\
  \begin{minipage}{14.2cm}
      \begin{minipage}[b]{7cm}
      \begin{center}
        \scalebox{0.45}{\input{figures/State_graphEcoliCoarse.pstex_t}}\\
        \hspace*{-.7cm}\mbox{(d)}
      \end{center}
      \end{minipage}\hfill
      \begin{minipage}[b]{7cm}
      \begin{center}
        \scalebox{0.45}{\input{figures/State_graphEcoliFine.pstex_t}}\\
        \hspace*{-.7cm}\mbox{(e)}
      \end{center}
      \end{minipage}
  \end{minipage}
\caption{\label{fig:example-refinement}  (a) Two-dimensional projection of a slice of the phase space of the \textit{E. coli} stress response model for the variables $x_{\crp}$ and $x_{\cya}$. (b)-(c) Partitioning into (b) mode domains and (c) flow domains of the projection. (d)-(e) Excerpts of the state transition graph resulting from the qualitative abstraction based on (d) mode domains and (e) flow domains.}
\end{figure*}

The results of this paper extend our previous work on the qualitative analysis of PADE models of genetic regulatory networks \cite{HdJ2039,HdJ2125}. In particular, we have defined a refined partitioning of the state space which underlies a qualitative abstraction preserving stronger properties of the qualitative dynamics of the system, \textit{i.e.} the derivative sign pattern. The resulting qualitative PA transition system is better adapted to the abstraction level of the experimental data, in the sense that it avoids verification of dynamical properties to be over-conservative. Consider Figure~\ref{fig:example-refinement}, which compares two-dimensional projections of a state-space slice of the stress response model. Depending on whether mode domains or flow domains are used as the abstraction criterion, the state transition graph will be different (compare (d) and (e) of Figure~\ref{fig:example-refinement}). Whereas the CTL formula $EF (\dot{x}_{\crp}>0 \ \wedge EF \ (\dot{x}_{\crp}<0))$ holds for the graph in (d), this is not true in (e), thus revealing that the coarse-grained abstraction may cause models to escape refutation by available experimental data. Judging from our experience with several PADE models of bacterial regulatory networks, the use of a fine-grained abstractions leads to only a modest increase in the size of the state transition graph \cite{HdJ2378}. That is, the increase in precision does not exclude the application of the refined abstraction to larger systems.

The hybrid character of the dynamics of genetic regulatory networks has stimulated the interest in the application of hybrid-systems methods and tools over the past few years \cite{HdJ2527,HdJ2536,HdJ2039,HdJ2354,HdJ2476}. Our approach differs from this related work on several counts. Whereas we use piecewise-affine deterministic models, other groups have employed multi-affine and related deterministic models \cite{HdJ2536,HdJ1032} or stochastic models \cite{HdJ2476}. These models are less restrictive and thus provide a more precise description of the network interactions. However, they are more difficult to analyze and in higher dimensions usually require the application of numerical techniques. This is not straightforward to achieve for most biological systems, since numerical information on parameter values is usually imprecise or simply not available.

The PADE models (\ref{eq:kinbasic}) in this paper have been well-studied in mathematical biology  \cite{HdJ2474,HdJ2516,HdJ2125,HdJ1619,HdJ1553,HdJ2515,HdJ1029,HdJ1030,HdJ1195,HdJ1899,HdJ1032,HdJ2488,HdJ1038,HdJ1039}, and have also formed the basis for other work in the field of hybrid systems \cite{HdJ2354}. However, contrary to \cite{HdJ2354}, we take into account the dynamics of the system on threshold hyperplanes, where equilibrium points and other phenomena of interest may occur \cite{HdJ2474,HdJ1899} (but see \cite{HdJ2527} for ideas on how to extend the approach in \cite{HdJ2354}). In \cite{HdJ2354} the partition of the state space is dynamically refined, so as to arrive at a qualitative PA transition system that is a better approximation of the original PA system. This requires the use of quantifier elimination methods \cite{HdJ2535} which allow to decide more general and more powerful properties than the rules proposed in Section~\ref{sec:computation} of this paper, but that also incur higher computational costs. The approach of this paper allows us to fully exploit the favorable mathematical properties of the PADE models (\ref{eq:kinbasic}), and thus promote the upscalability of the method to large and complex networks (Section~\ref{sec:application}).

From a more general perspective, our approach can be seen as an application of the notion of discrete abstraction, commonly used to study the dynamics of systems with an infinite number of states \cite{HdJ2540,HdJ2529,HdJ2205,HdJ2370,HdJ2530,HdJ1908}. Much work has focused on the identification of classes of continuous and hybrid dynamical systems for which bisimulation relations with finite transition systems are guaranteed to exist. The results of this paper can be seen as showing that the weaker simulation relation may also be of considerable practical interest, especially for classes of systems for which the existence of a finite bisimulation cannot be guaranteed. Discrete abstraction criteria similar to the one used in this paper, based on the sign of the (higher) derivatives of continuous variables, have been proposed by other authors in the fields of hybrids systems \cite{HdJ2540,HdJ1908} and control theory and qualitative reasoning \cite{HdJ1929,HdJ1881,HdJ619,HdJ2030}. In comparison with these approaches, our work deals with a less general class of models. However, this allows the development and implementation of efficient, tailored algorithms for the practical computation of the qualitative dynamics of the system, even on (intersections of) threshold hyperplanes, where discontinuities may occur.

The possibility to use efficient algorithms for the computation of the qualitative PA transition system rests, to a large extent, on the approximation of the set $K(x)$ in (\ref{eq:kininclusion_f}) by the set $H(x)$ in (\ref{eq:kininclusion}). Because the latter set is hyperrectangular, the computation of domains, transitions, and sign patterns can be carried out seperately in every dimension, using the ordering of parameter values fixed by inequality constraints. Because $H(x)$ is an overapproximation of $K(x)$, the state transition graph may contain sequences of states that would not occur in the graph obtained by using $K(x)$. As a consequence, a PADE model may fail to be rejected by an observed time-series of measurements of the concentration variables. However, due to the fact that the approximation of $H(x)$ by $K(x)$ is conservative, a PADE model will never be falsely rejected. An obvious direction for further research would be to see whether finer overapproximations of $K(x)$ can be found that still allow tailored symbolic algorithms to be used that do not compromise the upscalability of the method.

\mbox{}\\
\noindent\textbf{Acknowledgments} The authors acknowledge financial support from the ARC initiative at INRIA (GDyn project), the ACI IMPBio initiative of the French Ministry for Research (BacAttract project), and the NEST programme of the European Commission (Hygeia project, NEST 4995). Jean-Luc Gouz\'{e} and Tewfik Sari have provided helpful comments on previous versions of this paper.

\appendix

\section{Proofs}
\label{app:proofs}
\setcounter{pr}{0}
\setcounter{tm}{0}

\begin{pr}[Reachability equivalence]\rm
For all $x, x' \in \Omega$, there exists a solution $\xi(t) \in \Sigma$ and $\tau, \tau'$, such that $0 \leq \tau \leq \tau'$, $\xi(\tau)=x$, and $\xi(\tau')=x'$ if and only if there exists a run $(x^0, \ldots, x^m)$ of  $\Sigma\mbox{-}\mathrm{TS}$ such that $x^0=x$ and  $x^m=x'$.
\label{tm:reach_equiv}
\end{pr}

\begin{proof}
We first prove necessity. Consider two points $x, x' \in \Omega$, a solution $\xi(t) \in \Xi_\Sigma$, and $\tau, \tau'$ such that $0 \leq \tau \leq \tau'$, $\xi(\tau)=x$, and $\xi(\tau')=x'$. If $\tau= \tau'$, then $(\xi(\tau))$ is a trivial run satisfying the conditions of the proposition. Otherwise, $\tau< \tau'$ and we denote by $D^0, \ldots, D^m$ the time-ordered sequence of flow domains traversed by $\xi(t)$ on the time interval $[\tau, \tau']$. $D^0, \ldots, D^m$ is a finite sequence, since by definition any solution of $\Sigma$ reaches and leaves finitely-many times a threshold hyperplane during a time interval. If $m=0$, then by Definition~\ref{df:concreteTS} there exists an $int$ transition from $\xi(\tau)$ to $\xi(\tau')$, and consequently, $(\xi(\tau), \xi(\tau'))$ is a run satisfying the conditions of the proposition. Otherwise, $m>0$, and we denote by $\sigma^0, \ldots, \sigma^{m-1}$, the switching times, formally defined as $\sigma^j=\sup\, \{t \in [\tau, \tau'] \mid \xi(t)\in D^{j}\}$, $j \in \{0, \ldots , m-1\}$. Finally, we introduce a sequence of time instants $\tau^0, \ldots, \tau^m$ such that $\xi(\tau^j)\in D^j$, $j\in \{ 0, \ldots ,m\}$. More precisely, we define $\tau^0$ as $\tau$, $\tau^j$ as $(\sigma^{j-1}+\sigma^j)/2$, for all $j \in \{ 1, \ldots ,m-1\}$, and $\tau^m$ as $\tau'$. It is not difficult to show by induction on $j$ that $(\xi(\tau^0), \ldots, \xi(\tau^j))$ is a run, for every $j\in \{ 0, \ldots ,m\}$ \cite{HdJ2538}.

Next, we prove sufficiency. Consider a run $(x^0, \ldots, x^m)$ of  $\Sigma\mbox{-}\mathrm{TS}$, with $x^0=x$ and  $x^m=x'$. If $m=0$, then $x=x'$ and any solution $\xi(t) \in \Xi_\Sigma$ with $\xi(0)=x$ satisfies the conditions of the proposition for $\tau=\tau'=0$. In the sequel, we suppose $m>0$. Then, by Definition~\ref{df:concreteTS}, there exists a sequence of solutions $(\xi^0(t), \ldots , \xi^{m-1}(t))$, and of time instants $(\tau^0, \ldots ,\tau^{m})$, such that for all $j \in \{ 0, \ldots m-1 \}$, $\xi^j(t)$ is defined on the time interval $[\tau^j, \tau^{j+1}]$, with $\tau^j<\tau^{j+1}$, and satisfies $\xi^j(\tau^j) = x^{j}$ and $\xi^j(\tau^{j+1})= x^{j+1}$. It can be straightforwardly shown \cite{HdJ2538} that the concatenation of the solutions $\xi^j(t)$, $j\in \{ 0, \ldots ,m-1\}$, is a solution that satisfies the conditions of the proposition.
\end{proof}

\setcounter{pr}{7}

\begin{pr}[Computation of $dim^-$ transition] \rm
\label{tm:qual_tr_b}
Let $D,D' \in \mathcal{D}$ and $D' \subseteq \partial D$. \linebreak $D \stackrel{dim^-}{\longrightarrow}_{\sim_\Omega} D'$ if and only if $\Psi(\mode(D)) \not = \{\}$ and there exist $x\in D$, $x'\in D'$, and $\psi\in \Psi(\mode(D))$, such that either (a) for all $i\in \{ 1, \ldots ,n\}$ for which $D'_i \subseteq \partial D_i$, it holds that
\begin{equation}
(\psi_i- x_i')(x_i'-x_i)>0,
\end{equation}
or (b) it holds that $\psi= x'$.
\end{pr}

\begin{proof}
Let $M = \mode(D)$. We first prove necessity by contraposition. If $\Psi(M) = \{\}$, then there exists no solution remaining in $D$ for some time, and a transition from $D$ to $D'$ is not possible. Otherwise, suppose that for all $x\in D$, $x'\in D'$, and $\psi \in \Psi(M)$, there exists some $i\in \{ 1, \ldots ,n\}$ for which $D'_i \subseteq \partial D_i$, $(\psi_i- x'_i)(x_i'-x_i) \leq 0$, and $\psi \not = x'$. We assume $x_i'-x_i>0$ for all $x\in D$, $x'\in D'$ (the case $x_i'-x_i<0$ goes analogously).

If the inequality is strict, then $\psi_i < x'_i$, for all $x'\in D'$ and $\psi \in \Psi(M)$. By Lemma~\ref{lem:monotone-singular}, for all solutions $\xi (t) \in \Xi_\Sigma$ in $D$, $\xi_i (t)$ monotonically converges towards $\Psi_i(M)$. As a consequence, no solution enters $D'$ from $D$, and there does not exist a transition $D \stackrel{dim^-}{\longrightarrow}_{\sim_\Omega} D'$. If the inequality is not strict, then by definition of the flow domain partition (Definition~\ref{df:flow_domain}), we have $\psi_i = x'_i$ for all $x' \in D'$ and $\psi \in \Psi(M)$. From the same definition it also follows that either $\psi_i = \psi_i(M)$ (if $M$ is regular) or $\psi_i = \max_{M' \in R(M)} \psi_i(M')$ (if $M$ is singular). It can be shown by construction in the way of the proof of Lemma~\ref{lem:monotone-singular-existence} that solutions reach  $D'$ only asymptotically, as $t \rightarrow \infty$. Moreover, from the definition of $\Psi(M)$ and Lemmas~\ref{lem:monotone-regular} and \ref{lem:monotone-singular}, it follows for every $j\in \{1, \ldots ,n\}$ that either $\xi_j(t) \in \Psi_j(M)$ or that $\xi_j(t)$ monotonically converges towards $\Psi_j(M)$, as long as $\xi(t)$ remains in $D$ (and thus in $M$). Therefore, because $\xi(t)$ reaches $x' \in D'$ asymptotically, we have $x' \in \Psi(M)$. This contradicts the assumption $\psi \not = x'$, so there does not exist a $dim^-$ transition.

Next, we prove sufficiency. Suppose $\Psi(M) \not = \{\}$ and there exist $x\in D$, $x'\in D'$, and $\psi\in \Psi(M)$, such that (a) for all $i\in \{1, \ldots ,n\}$ for which $D_i \subseteq \partial D'_i$, it holds that $(\psi_i- x'_i)(x_i'-x_i)>0$, or (b) it holds that $\psi= x'$. Consider (a) and assume $x_i'-x_i>0$ for all $x\in D$, $x'\in D'$ (the case $x_i'-x_i<0$ goes analogously). As a consequence, $x_i < x'_i < \psi_i$ for all $x\in D$ and $x'\in D'$, and some $\psi \in \Psi(M)$. By Lemma~\ref{lem:monotone-singular-existence}, for all $x \in D$ there exists a solution $\xi (t) \in \Xi_\Sigma$ in $D$, with $\xi (0) = x$, which monotonically converges towards $\psi$. We choose $x$ such that the corresponding solution enters $D'$ from $D$, and thus obtain a transition $D \stackrel{dim^-}{\longrightarrow}_{\sim_\Omega} D'$. In case (b) we find $x_i < x'_i = \psi_i$, and by the same argument there exists a solution that enters $D'$ from $D$. However, in this case the solution only reaches $D'$ asymptotically, as $t \rightarrow \infty$ (see the proof of necessity).
\end{proof}

\bibliography{biblio}
\bibliographystyle{plain}
\newpage
\tableofcontents
\end{document}

%% file: figures/exnetwork.pstex_t
\begin{picture}(0,0)%
\includegraphics{exnetwork.pstex}%
\end{picture}%
\setlength{\unitlength}{3947sp}%
\begingroup\makeatletter\ifx\SetFigFont\undefined%
\gdef\SetFigFont#1#2#3#4#5{%
  \reset@font\fontsize{#1}{#2pt}%
  \fontfamily{#3}\fontseries{#4}\fontshape{#5}%
  \selectfont}%
\fi\endgroup%
\begin{picture}(6794,2220)(2979,-5611)
\end{picture}%

%% file: figures/exbox1_coarse.pstex_t
\begin{picture}(0,0)%
\includegraphics{exbox1_coarse.pstex}%
\end{picture}%
\setlength{\unitlength}{3947sp}%
\begingroup\makeatletter\ifx\SetFigFont\undefined%
\gdef\SetFigFont#1#2#3#4#5{%
  \reset@font\fontsize{#1}{#2pt}%
  \fontfamily{#3}\fontseries{#4}\fontshape{#5}%
  \selectfont}%
\fi\endgroup%
\begin{picture}(5197,3403)(4126,-4417)
\put(4126,-1711){\makebox(0,0)[lb]{\smash{\SetFigFont{17}{20.4}{\rmdefault}{\mddefault}{\updefault}$\mathit{max}_b$}}}
\put(4651,-4261){\makebox(0,0)[lb]{\smash{\SetFigFont{17}{20.4}{\rmdefault}{\mddefault}{\updefault}$0$}}}
\put(4126,-1261){\makebox(0,0)[lb]{\smash{\SetFigFont{17}{20.4}{\rmdefault}{\mddefault}{\updefault}$x_b$}}}
\put(8101,-4261){\makebox(0,0)[lb]{\smash{\SetFigFont{17}{20.4}{\rmdefault}{\mddefault}{\updefault}$\mathit{max}_a$}}}
\put(8926,-4336){\makebox(0,0)[lb]{\smash{\SetFigFont{17}{20.4}{\rmdefault}{\mddefault}{\updefault}$x_a$}}}
\put(5926,-4261){\makebox(0,0)[lb]{\smash{\SetFigFont{17}{20.4}{\rmdefault}{\mddefault}{\updefault}$\theta_a^1$}}}
\put(7126,-4261){\makebox(0,0)[lb]{\smash{\SetFigFont{17}{20.4}{\rmdefault}{\mddefault}{\updefault}$\theta_a^2$}}}
\put(5176,-3361){\makebox(0,0)[lb]{\smash{\SetFigFont{17}{20.4}{\rmdefault}{\mddefault}{\updefault}$M^1$}}}
\put(5851,-3361){\makebox(0,0)[lb]{\smash{\SetFigFont{17}{20.4}{\rmdefault}{\mddefault}{\updefault}$M^2$}}}
\put(7051,-3361){\makebox(0,0)[lb]{\smash{\SetFigFont{17}{20.4}{\rmdefault}{\mddefault}{\updefault}$M^{4}$}}}
\put(7576,-3361){\makebox(0,0)[lb]{\smash{\SetFigFont{17}{20.4}{\rmdefault}{\mddefault}{\updefault}$M^{5}$}}}
\put(5176,-2836){\makebox(0,0)[lb]{\smash{\SetFigFont{17}{20.4}{\rmdefault}{\mddefault}{\updefault}$M^{6}$}}}
\put(5851,-2836){\makebox(0,0)[lb]{\smash{\SetFigFont{17}{20.4}{\rmdefault}{\mddefault}{\updefault}$M^{7}$}}}
\put(6376,-2836){\makebox(0,0)[lb]{\smash{\SetFigFont{17}{20.4}{\rmdefault}{\mddefault}{\updefault}$M^{8}$}}}
\put(7051,-2836){\makebox(0,0)[lb]{\smash{\SetFigFont{17}{20.4}{\rmdefault}{\mddefault}{\updefault}$M^{9}$}}}
\put(5176,-2161){\makebox(0,0)[lb]{\smash{\SetFigFont{17}{20.4}{\rmdefault}{\mddefault}{\updefault}$M^{11}$}}}
\put(5851,-2161){\makebox(0,0)[lb]{\smash{\SetFigFont{17}{20.4}{\rmdefault}{\mddefault}{\updefault}$M^{12}$}}}
\put(6376,-2161){\makebox(0,0)[lb]{\smash{\SetFigFont{17}{20.4}{\rmdefault}{\mddefault}{\updefault}$M^{13}$}}}
\put(7051,-2161){\makebox(0,0)[lb]{\smash{\SetFigFont{17}{20.4}{\rmdefault}{\mddefault}{\updefault}$M^{14}$}}}
\put(7576,-2161){\makebox(0,0)[lb]{\smash{\SetFigFont{17}{20.4}{\rmdefault}{\mddefault}{\updefault}$M^{15}$}}}
\put(7576,-2836){\makebox(0,0)[lb]{\smash{\SetFigFont{17}{20.4}{\rmdefault}{\mddefault}{\updefault}$M^{10}$}}}
\put(6376,-3361){\makebox(0,0)[lb]{\smash{\SetFigFont{17}{20.4}{\rmdefault}{\mddefault}{\updefault}$M^{3}$}}}
\put(4276,-2836){\makebox(0,0)[lb]{\smash{\SetFigFont{17}{20.4}{\rmdefault}{\mddefault}{\updefault}$\theta_b$}}}
\end{picture}

%% file: figures/exbox2_coarse.pstex_t
\begin{picture}(0,0)%
\includegraphics{exbox2_coarse.pstex}%
\end{picture}%
\setlength{\unitlength}{3947sp}%
\begingroup\makeatletter\ifx\SetFigFont\undefined%
\gdef\SetFigFont#1#2#3#4#5{%
  \reset@font\fontsize{#1}{#2pt}%
  \fontfamily{#3}\fontseries{#4}\fontshape{#5}%
  \selectfont}%
\fi\endgroup%
\begin{picture}(5851,3328)(3751,-4342)
\put(8926,-4261){\makebox(0,0)[lb]{\smash{{\SetFigFont{17}{20.4}{\rmdefault}{\mddefault}{\updefault}$x_a$}}}}
\put(4126,-1336){\makebox(0,0)[lb]{\smash{{\SetFigFont{17}{20.4}{\rmdefault}{\mddefault}{\updefault}$x_b$}}}}
\put(5251,-3436){\makebox(0,0)[lb]{\smash{{\SetFigFont{17}{20.4}{\rmdefault}{\mddefault}{\updefault}$M^{1}$}}}}
\put(6376,-3436){\makebox(0,0)[lb]{\smash{{\SetFigFont{17}{20.4}{\rmdefault}{\mddefault}{\updefault}$M^{3}$}}}}
\put(7051,-3136){\makebox(0,0)[lb]{\smash{{\SetFigFont{17}{20.4}{\rmdefault}{\mddefault}{\updefault}$M^{4}$}}}}
\put(7576,-3361){\makebox(0,0)[lb]{\smash{{\SetFigFont{17}{20.4}{\rmdefault}{\mddefault}{\updefault}$M^{5}$}}}}
\put(5851,-3136){\makebox(0,0)[lb]{\smash{{\SetFigFont{17}{20.4}{\rmdefault}{\mddefault}{\updefault}$M^{2}$}}}}
\put(5326,-2589){\makebox(0,0)[lb]{\smash{{\SetFigFont{17}{20.4}{\rmdefault}{\mddefault}{\updefault}$M^{11}$}}}}
\put(7876,-1411){\makebox(0,0)[lb]{\smash{{\SetFigFont{17}{20.4}{\rmdefault}{\mddefault}{\updefault}$\kappa_a/\gamma_a$}}}}
\put(8551,-2386){\makebox(0,0)[lb]{\smash{{\SetFigFont{17}{20.4}{\rmdefault}{\mddefault}{\updefault}$\kappa_b/\gamma_b$}}}}
\put(4576,-4261){\makebox(0,0)[lb]{\smash{{\SetFigFont{17}{20.4}{\rmdefault}{\mddefault}{\updefault}$\Psi(M^{5})$}}}}
\put(3751,-2386){\makebox(0,0)[lb]{\smash{{\SetFigFont{17}{20.4}{\rmdefault}{\mddefault}{\updefault}$\Psi(M^{11})$}}}}
\put(7801,-4261){\makebox(0,0)[lb]{\smash{{\SetFigFont{17}{20.4}{\rmdefault}{\mddefault}{\updefault}$\Psi(M^{3})$}}}}
\put(6826,-4261){\makebox(0,0)[lb]{\smash{{\SetFigFont{17}{20.4}{\rmdefault}{\mddefault}{\updefault}$\Psi(M^{4})$}}}}
\put(7276,-2236){\makebox(0,0)[lb]{\smash{{\SetFigFont{17}{20.4}{\rmdefault}{\mddefault}{\updefault}$\Psi(M^{1})$}}}}
\end{picture}%

%% file: figures/exbox1_fine.pstex_t
\begin{picture}(0,0)%
\includegraphics{exbox1_fine.pstex}%
\end{picture}%
\setlength{\unitlength}{3947sp}%
\begingroup\makeatletter\ifx\SetFigFont\undefined%
\gdef\SetFigFont#1#2#3#4#5{%
  \reset@font\fontsize{#1}{#2pt}%
  \fontfamily{#3}\fontseries{#4}\fontshape{#5}%
  \selectfont}%
\fi\endgroup%
\begin{picture}(5272,3403)(4051,-4417)
\put(8176,-4261){\makebox(0,0)[lb]{\smash{\SetFigFont{17}{20.4}{\rmdefault}{\mddefault}{\updefault}$\mathit{max}_a$}}}
\put(4126,-1711){\makebox(0,0)[lb]{\smash{\SetFigFont{17}{20.4}{\rmdefault}{\mddefault}{\updefault}$\mathit{max}_b$}}}
\put(4651,-4261){\makebox(0,0)[lb]{\smash{\SetFigFont{17}{20.4}{\rmdefault}{\mddefault}{\updefault}$0$}}}
\put(4276,-2836){\makebox(0,0)[lb]{\smash{\SetFigFont{17}{20.4}{\rmdefault}{\mddefault}{\updefault}$\theta_b$}}}
\put(4051,-2311){\makebox(0,0)[lb]{\smash{\SetFigFont{17}{20.4}{\rmdefault}{\mddefault}{\updefault}$\kappa_b/\gamma_b$}}}
\put(4276,-1336){\makebox(0,0)[lb]{\smash{\SetFigFont{17}{20.4}{\rmdefault}{\mddefault}{\updefault}$x_b$}}}
\put(8926,-4336){\makebox(0,0)[lb]{\smash{\SetFigFont{17}{20.4}{\rmdefault}{\mddefault}{\updefault}$x_a$}}}
\put(5926,-4261){\makebox(0,0)[lb]{\smash{\SetFigFont{17}{20.4}{\rmdefault}{\mddefault}{\updefault}$\theta_a^1$}}}
\put(7051,-4261){\makebox(0,0)[lb]{\smash{\SetFigFont{17}{20.4}{\rmdefault}{\mddefault}{\updefault}$\theta_a^2$}}}
\put(5926,-2611){\makebox(0,0)[lb]{\smash{\SetFigFont{14}{16.8}{\rmdefault}{\mddefault}{\updefault}$D^{12.1}$}}}
\put(5926,-2386){\makebox(0,0)[lb]{\smash{\SetFigFont{14}{16.8}{\rmdefault}{\mddefault}{\updefault}$D^{12.2}$}}}
\put(5926,-1936){\makebox(0,0)[lb]{\smash{\SetFigFont{14}{16.8}{\rmdefault}{\mddefault}{\updefault}$D^{12.3}$}}}
\put(5326,-1936){\makebox(0,0)[lb]{\smash{\SetFigFont{14}{16.8}{\rmdefault}{\mddefault}{\updefault}$D^{11.6}$}}}
\put(4726,-1936){\makebox(0,0)[lb]{\smash{\SetFigFont{14}{16.8}{\rmdefault}{\mddefault}{\updefault}$D^{11.5}$}}}
\put(5326,-2386){\makebox(0,0)[lb]{\smash{\SetFigFont{14}{16.8}{\rmdefault}{\mddefault}{\updefault}$D^{11.4}$}}}
\put(4726,-2386){\makebox(0,0)[lb]{\smash{\SetFigFont{14}{16.8}{\rmdefault}{\mddefault}{\updefault}$D^{11.3}$}}}
\put(5326,-2611){\makebox(0,0)[lb]{\smash{\SetFigFont{14}{16.8}{\rmdefault}{\mddefault}{\updefault}$D^{11.2}$}}}
\put(4726,-2611){\makebox(0,0)[lb]{\smash{\SetFigFont{14}{16.8}{\rmdefault}{\mddefault}{\updefault}$D^{11.1}$}}}
\put(7126,-4036){\makebox(0,0)[lb]{\smash{\SetFigFont{14}{16.8}{\rmdefault}{\mddefault}{\updefault}$D^{4.1}$}}}
\put(7126,-3436){\makebox(0,0)[lb]{\smash{\SetFigFont{14}{16.8}{\rmdefault}{\mddefault}{\updefault}$D^{4.2}$}}}
\put(7126,-2236){\makebox(0,0)[lb]{\smash{\SetFigFont{14}{16.8}{\rmdefault}{\mddefault}{\updefault}$D^{14.1}$}}}
\put(7726,-2236){\makebox(0,0)[lb]{\smash{\SetFigFont{14}{16.8}{\rmdefault}{\mddefault}{\updefault}$D^{15.1}$}}}
\put(6526,-2236){\makebox(0,0)[lb]{\smash{\SetFigFont{14}{16.8}{\rmdefault}{\mddefault}{\updefault}$D^{13.1}$}}}
\put(7726,-3436){\makebox(0,0)[lb]{\smash{\SetFigFont{14}{16.8}{\rmdefault}{\mddefault}{\updefault}$D^{5.2}$}}}
\put(5326,-2836){\makebox(0,0)[lb]{\smash{\SetFigFont{14}{16.8}{\rmdefault}{\mddefault}{\updefault}$D^{6.2}$}}}
\put(4726,-2836){\makebox(0,0)[lb]{\smash{\SetFigFont{14}{16.8}{\rmdefault}{\mddefault}{\updefault}$D^{6.1}$}}}
\put(5926,-2836){\makebox(0,0)[lb]{\smash{\SetFigFont{14}{16.8}{\rmdefault}{\mddefault}{\updefault}$D^{7.1}$}}}
\put(6451,-2836){\makebox(0,0)[lb]{\smash{\SetFigFont{14}{16.8}{\rmdefault}{\mddefault}{\updefault}$D^{8.1}$}}}
\put(7126,-2836){\makebox(0,0)[lb]{\smash{\SetFigFont{14}{16.8}{\rmdefault}{\mddefault}{\updefault}$D^{9.1}$}}}
\put(7726,-2836){\makebox(0,0)[lb]{\smash{\SetFigFont{14}{16.8}{\rmdefault}{\mddefault}{\updefault}$D^{10.1}$}}}
\put(6526,-3436){\makebox(0,0)[lb]{\smash{\SetFigFont{14}{16.8}{\rmdefault}{\mddefault}{\updefault}$D^{3.2}$}}}
\put(5926,-3436){\makebox(0,0)[lb]{\smash{\SetFigFont{14}{16.8}{\rmdefault}{\mddefault}{\updefault}$D^{2.2}$}}}
\put(5326,-3436){\makebox(0,0)[lb]{\smash{\SetFigFont{14}{16.8}{\rmdefault}{\mddefault}{\updefault}$D^{1.1}$}}}
\put(5926,-4036){\makebox(0,0)[lb]{\smash{\SetFigFont{14}{16.8}{\rmdefault}{\mddefault}{\updefault}$D^{2.1}$}}}
\put(6526,-4036){\makebox(0,0)[lb]{\smash{\SetFigFont{14}{16.8}{\rmdefault}{\mddefault}{\updefault}$D^{3.1}$}}}
\put(7726,-4036){\makebox(0,0)[lb]{\smash{\SetFigFont{14}{16.8}{\rmdefault}{\mddefault}{\updefault}$D^{5.1}$}}}
\end{picture}

%% file: figures/statetrgraph_hs.pstex_t
\begin{picture}(0,0)%
\includegraphics{statetrgraph_hs.pstex}%
\end{picture}%
\setlength{\unitlength}{3947sp}%
\begingroup\makeatletter\ifx\SetFigFont\undefined%
\gdef\SetFigFont#1#2#3#4#5{%
  \reset@font\fontsize{#1}{#2pt}%
  \fontfamily{#3}\fontseries{#4}\fontshape{#5}%
  \selectfont}%
\fi\endgroup%
\begin{picture}(6825,4939)(4126,-6794)
\put(7051,-3211){\makebox(0,0)[lb]{\smash{\SetFigFont{12}{14.4}{\rmdefault}{\mddefault}{\updefault}$D^{12.1}$}}}
\put(7051,-2611){\makebox(0,0)[lb]{\smash{\SetFigFont{12}{14.4}{\rmdefault}{\mddefault}{\updefault}$D^{12.2}$}}}
\put(7051,-2011){\makebox(0,0)[lb]{\smash{\SetFigFont{12}{14.4}{\rmdefault}{\mddefault}{\updefault}$D^{12.3}$}}}
\put(8251,-6661){\makebox(0,0)[lb]{\smash{\SetFigFont{12}{14.4}{\rmdefault}{\mddefault}{\updefault}$D^{3.1}$}}}
\put(10651,-6661){\makebox(0,0)[lb]{\smash{\SetFigFont{12}{14.4}{\rmdefault}{\mddefault}{\updefault}$D^{5.1}$}}}
\put(10651,-5461){\makebox(0,0)[lb]{\smash{\SetFigFont{12}{14.4}{\rmdefault}{\mddefault}{\updefault}$D^{5.2}$}}}
\put(10951,-3961){\makebox(0,0)[lb]{\smash{\SetFigFont{12}{14.4}{\rmdefault}{\mddefault}{\updefault}$D^{10.1}$}}}
\put(9451,-2611){\makebox(0,0)[lb]{\smash{\SetFigFont{12}{14.4}{\rmdefault}{\mddefault}{\updefault}$D^{14.1}$}}}
\put(8251,-5461){\makebox(0,0)[lb]{\smash{\SetFigFont{12}{14.4}{\rmdefault}{\mddefault}{\updefault}$D^{3.2}$}}}
\put(8551,-3961){\makebox(0,0)[lb]{\smash{\SetFigFont{12}{14.4}{\rmdefault}{\mddefault}{\updefault}$D^{8.1}$}}}
\put(7051,-6661){\makebox(0,0)[lb]{\smash{\SetFigFont{12}{14.4}{\rmdefault}{\mddefault}{\updefault}$D^{2.1}$}}}
\put(7051,-5461){\makebox(0,0)[lb]{\smash{\SetFigFont{12}{14.4}{\rmdefault}{\mddefault}{\updefault}$D^{2.2}$}}}
\put(5326,-5311){\makebox(0,0)[lb]{\smash{\SetFigFont{12}{14.4}{\rmdefault}{\mddefault}{\updefault}$D^{1.1}$}}}
\put(9751,-5461){\makebox(0,0)[lb]{\smash{\SetFigFont{12}{14.4}{\rmdefault}{\mddefault}{\updefault}$D^{4.2}$}}}
\put(4126,-2761){\makebox(0,0)[lb]{\smash{\SetFigFont{12}{14.4}{\rmdefault}{\mddefault}{\updefault}$D^{11.3}$}}}
\put(9451,-6736){\makebox(0,0)[lb]{\smash{\SetFigFont{12}{14.4}{\rmdefault}{\mddefault}{\updefault}$D^{4.1}$}}}
\put(10726,-2611){\makebox(0,0)[lb]{\smash{\SetFigFont{12}{14.4}{\rmdefault}{\mddefault}{\updefault}$D^{15.1}$}}}
\put(8251,-2536){\makebox(0,0)[lb]{\smash{\SetFigFont{12}{14.4}{\rmdefault}{\mddefault}{\updefault}$D^{13.1}$}}}
\put(5926,-2011){\makebox(0,0)[lb]{\smash{\SetFigFont{12}{14.4}{\rmdefault}{\mddefault}{\updefault}$D^{11.6}$}}}
\put(5926,-2611){\makebox(0,0)[lb]{\smash{\SetFigFont{12}{14.4}{\rmdefault}{\mddefault}{\updefault}$D^{11.4}$}}}
\put(5926,-3211){\makebox(0,0)[lb]{\smash{\SetFigFont{12}{14.4}{\rmdefault}{\mddefault}{\updefault}$D^{11.2}$}}}
\put(4126,-3361){\makebox(0,0)[lb]{\smash{\SetFigFont{12}{14.4}{\rmdefault}{\mddefault}{\updefault}$D^{11.1}$}}}
\put(4126,-2161){\makebox(0,0)[lb]{\smash{\SetFigFont{12}{14.4}{\rmdefault}{\mddefault}{\updefault}$D^{11.5}$}}}
\put(4126,-3961){\makebox(0,0)[lb]{\smash{\SetFigFont{12}{14.4}{\rmdefault}{\mddefault}{\updefault}$D^{6.1}$}}}
\put(7426,-3961){\makebox(0,0)[lb]{\smash{\SetFigFont{12}{14.4}{\rmdefault}{\mddefault}{\updefault}$D^{7.1}$}}}
\put(5401,-4036){\makebox(0,0)[lb]{\smash{\SetFigFont{12}{14.4}{\rmdefault}{\mddefault}{\updefault}$D^{6.2}$}}}
\put(9751,-4036){\makebox(0,0)[lb]{\smash{\SetFigFont{12}{14.4}{\rmdefault}{\mddefault}{\updefault}$D^{9.1}$}}}
\end{picture}

%% file: figures/sign_M1.pstex_t
\begin{picture}(0,0)%
\includegraphics{sign_M1.pstex}%
\end{picture}%
\setlength{\unitlength}{3947sp}%
\begingroup\makeatletter\ifx\SetFigFont\undefined%
\gdef\SetFigFont#1#2#3#4#5{%
  \reset@font\fontsize{#1}{#2pt}%
  \fontfamily{#3}\fontseries{#4}\fontshape{#5}%
  \selectfont}%
\fi\endgroup%
\begin{picture}(5272,3553)(4051,-4567)
\put(4126,-1711){\makebox(0,0)[lb]{\smash{\SetFigFont{17}{20.4}{\rmdefault}{\mddefault}{\updefault}$\mathit{max}_b$}}}
\put(4651,-4261){\makebox(0,0)[lb]{\smash{\SetFigFont{17}{20.4}{\rmdefault}{\mddefault}{\updefault}$0$}}}
\put(4276,-2836){\makebox(0,0)[lb]{\smash{\SetFigFont{17}{20.4}{\rmdefault}{\mddefault}{\updefault}$\theta_b$}}}
\put(4051,-2311){\makebox(0,0)[lb]{\smash{\SetFigFont{17}{20.4}{\rmdefault}{\mddefault}{\updefault}$\kappa_b/\gamma_b$}}}
\put(4126,-1336){\makebox(0,0)[lb]{\smash{\SetFigFont{17}{20.4}{\rmdefault}{\mddefault}{\updefault}$x_b$}}}
\put(5926,-4261){\makebox(0,0)[lb]{\smash{\SetFigFont{17}{20.4}{\rmdefault}{\mddefault}{\updefault}$\theta_a^1$}}}
\put(7126,-4261){\makebox(0,0)[lb]{\smash{\SetFigFont{17}{20.4}{\rmdefault}{\mddefault}{\updefault}$\theta_a^2$}}}
\put(8926,-4336){\makebox(0,0)[lb]{\smash{\SetFigFont{17}{20.4}{\rmdefault}{\mddefault}{\updefault}$x_a$}}}
\put(8101,-4261){\makebox(0,0)[lb]{\smash{\SetFigFont{17}{20.4}{\rmdefault}{\mddefault}{\updefault}$\mathit{max}_a$}}}
\put(5176,-3586){\makebox(0,0)[lb]{\smash{\SetFigFont{17}{20.4}{\rmdefault}{\mddefault}{\updefault}$D^{1.1}$}}}
\put(7651,-4486){\makebox(0,0)[lb]{\smash{\SetFigFont{17}{20.4}{\rmdefault}{\mddefault}{\updefault}$\kappa_a/\gamma_a$}}}
\put(7276,-2161){\makebox(0,0)[lb]{\smash{\SetFigFont{17}{20.4}{\rmdefault}{\mddefault}{\updefault}$\Psi(\mode(D^{1.1}))$}}}
\end{picture}

%% file: figures/sign_M11.pstex_t
\begin{picture}(0,0)%
\includegraphics{sign_M11.pstex}%
\end{picture}%
\setlength{\unitlength}{3947sp}%
\begingroup\makeatletter\ifx\SetFigFont\undefined%
\gdef\SetFigFont#1#2#3#4#5{%
  \reset@font\fontsize{#1}{#2pt}%
  \fontfamily{#3}\fontseries{#4}\fontshape{#5}%
  \selectfont}%
\fi\endgroup%
\begin{picture}(5551,3403)(4051,-4417)
\put(4126,-1711){\makebox(0,0)[lb]{\smash{{\SetFigFont{17}{20.4}{\rmdefault}{\mddefault}{\updefault}$\mathit{max}_b$}}}}
\put(4651,-4261){\makebox(0,0)[lb]{\smash{{\SetFigFont{17}{20.4}{\rmdefault}{\mddefault}{\updefault}$0$}}}}
\put(4276,-2836){\makebox(0,0)[lb]{\smash{{\SetFigFont{17}{20.4}{\rmdefault}{\mddefault}{\updefault}$\theta_b$}}}}
\put(4126,-1336){\makebox(0,0)[lb]{\smash{{\SetFigFont{17}{20.4}{\rmdefault}{\mddefault}{\updefault}$x_b$}}}}
\put(5926,-4261){\makebox(0,0)[lb]{\smash{{\SetFigFont{17}{20.4}{\rmdefault}{\mddefault}{\updefault}$\theta_a^1$}}}}
\put(7126,-4261){\makebox(0,0)[lb]{\smash{{\SetFigFont{17}{20.4}{\rmdefault}{\mddefault}{\updefault}$\theta_a^2$}}}}
\put(8926,-4336){\makebox(0,0)[lb]{\smash{{\SetFigFont{17}{20.4}{\rmdefault}{\mddefault}{\updefault}$x_a$}}}}
\put(8101,-4261){\makebox(0,0)[lb]{\smash{{\SetFigFont{17}{20.4}{\rmdefault}{\mddefault}{\updefault}$\mathit{max}_a$}}}}
\put(5851,-3586){\makebox(0,0)[lb]{\smash{{\SetFigFont{17}{20.4}{\rmdefault}{\mddefault}{\updefault}$D^{2.2}$}}}}
\put(7051,-3586){\makebox(0,0)[lb]{\smash{{\SetFigFont{17}{20.4}{\rmdefault}{\mddefault}{\updefault}$D^{4.2}$}}}}
\put(7276,-3886){\makebox(0,0)[lb]{\smash{{\SetFigFont{17}{20.4}{\rmdefault}{\mddefault}{\updefault}$\Psi(\mode(D^{4.2}))$}}}}
\put(4051,-2386){\makebox(0,0)[lb]{\smash{{\SetFigFont{17}{20.4}{\rmdefault}{\mddefault}{\updefault}$\kappa_b/\gamma_b$}}}}
\end{picture}%

%% file: figures/comp_trans_M3.pstex_t
\begin{picture}(0,0)%
\includegraphics{comp_trans_M3.pstex}%
\end{picture}%
\setlength{\unitlength}{3947sp}%
\begingroup\makeatletter\ifx\SetFigFont\undefined%
\gdef\SetFigFont#1#2#3#4#5{%
  \reset@font\fontsize{#1}{#2pt}%
  \fontfamily{#3}\fontseries{#4}\fontshape{#5}%
  \selectfont}%
\fi\endgroup%
\begin{picture}(5626,3403)(4126,-4417)
\put(9076,-4336){\makebox(0,0)[lb]{\smash{{\SetFigFont{17}{20.4}{\rmdefault}{\mddefault}{\updefault}$x_a$}}}}
\put(4126,-1711){\makebox(0,0)[lb]{\smash{{\SetFigFont{17}{20.4}{\rmdefault}{\mddefault}{\updefault}$\mathit{max}_b$}}}}
\put(4651,-4261){\makebox(0,0)[lb]{\smash{{\SetFigFont{17}{20.4}{\rmdefault}{\mddefault}{\updefault}$0$}}}}
\put(4276,-1336){\makebox(0,0)[lb]{\smash{{\SetFigFont{17}{20.4}{\rmdefault}{\mddefault}{\updefault}$x_b$}}}}
\put(5926,-4261){\makebox(0,0)[lb]{\smash{{\SetFigFont{17}{20.4}{\rmdefault}{\mddefault}{\updefault}$\theta_a^1$}}}}
\put(7126,-4261){\makebox(0,0)[lb]{\smash{{\SetFigFont{17}{20.4}{\rmdefault}{\mddefault}{\updefault}$\theta_a^2$}}}}
\put(4276,-2836){\makebox(0,0)[lb]{\smash{{\SetFigFont{17}{20.4}{\rmdefault}{\mddefault}{\updefault}$\theta_b$}}}}
\put(6601,-3136){\makebox(0,0)[lb]{\smash{{\SetFigFont{17}{20.4}{\rmdefault}{\mddefault}{\updefault}$D^{3.2}$}}}}
\put(6601,-4036){\makebox(0,0)[lb]{\smash{{\SetFigFont{17}{20.4}{\rmdefault}{\mddefault}{\updefault}$D^{3.1}$}}}}
\put(5851,-3586){\makebox(0,0)[lb]{\smash{{\SetFigFont{17}{20.4}{\rmdefault}{\mddefault}{\updefault}$D^{2.2}$}}}}
\put(7051,-3436){\makebox(0,0)[lb]{\smash{{\SetFigFont{17}{20.4}{\rmdefault}{\mddefault}{\updefault}$D^{4.2}$}}}}
\put(5851,-4036){\makebox(0,0)[lb]{\smash{{\SetFigFont{17}{20.4}{\rmdefault}{\mddefault}{\updefault}$D^{2.1}$}}}}
\put(8176,-4261){\makebox(0,0)[lb]{\smash{{\SetFigFont{17}{20.4}{\rmdefault}{\mddefault}{\updefault}$\mathit{max}_a$}}}}
\put(8026,-3811){\makebox(0,0)[lb]{\smash{{\SetFigFont{17}{20.4}{\rmdefault}{\mddefault}{\updefault}$\Psi(\mode(D^{3.2}))$}}}}
\put(7591,-1401){\makebox(0,0)[lb]{\smash{{\SetFigFont{17}{20.4}{\rmdefault}{\mddefault}{\updefault}$\kappa_a/\gamma_a$}}}}
\end{picture}%

%% file: figures/Ecoli_network.pstex_t
\begin{picture}(0,0)%
\includegraphics{Ecoli_network.pstex}%
\end{picture}%
\setlength{\unitlength}{4144sp}%
\begingroup\makeatletter\ifx\SetFigFont\undefined%
\gdef\SetFigFont#1#2#3#4#5{%
  \reset@font\fontsize{#1}{#2pt}%
  \fontfamily{#3}\fontseries{#4}\fontshape{#5}%
  \selectfont}%
\fi\endgroup%
\begin{picture}(11994,4299)(259,-7768)
\put(10253,-5981){\makebox(0,0)[lb]{\smash{{\SetFigFont{12}{14.4}{\rmdefault}{\mddefault}{\updefault}Synthesis of protein Fis }}}}
\put(10253,-6161){\makebox(0,0)[lb]{\smash{{\SetFigFont{12}{14.4}{\rmdefault}{\mddefault}{\updefault}from gene {\sl fis}}}}}
\put(2677,-4593){\makebox(0,0)[lb]{\smash{{\SetFigFont{12}{14.4}{\rmdefault}{\mddefault}{\updefault}GyrAB}}}}
\put(1441,-4921){\makebox(0,0)[lb]{\smash{{\SetFigFont{12}{14.4}{\rmdefault}{\mddefault}{\updefault}{\sl gyrAB}}}}}
\put(1486,-6766){\makebox(0,0)[lb]{\smash{{\SetFigFont{12}{14.4}{\rmdefault}{\mddefault}{\updefault}{\sl topA}}}}}
\put(7426,-5911){\makebox(0,0)[lb]{\smash{{\SetFigFont{12}{14.4}{\rmdefault}{\mddefault}{\updefault}{\sl crp}}}}}
\put(10576,-4966){\makebox(0,0)[lb]{\smash{{\SetFigFont{12}{14.4}{\rmdefault}{\mddefault}{\updefault}{\sl cya}}}}}
\put(7063,-6447){\makebox(0,0)[lb]{\smash{{\SetFigFont{12}{14.4}{\rmdefault}{\mddefault}{\updefault}stable RNAs}}}}
\put(7467,-4420){\makebox(0,0)[lb]{\smash{{\SetFigFont{9}{10.8}{\rmdefault}{\mddefault}{\updefault}cAMP$\cdot$CRP}}}}
\put(9614,-4500){\makebox(0,0)[lb]{\smash{{\SetFigFont{12}{14.4}{\rmdefault}{\mddefault}{\updefault}Cya}}}}
\put(5986,-6856){\makebox(0,0)[lb]{\smash{{\SetFigFont{12}{14.4}{\rmdefault}{\mddefault}{\updefault}{\sl rrn}}}}}
\put(5221,-5326){\makebox(0,0)[lb]{\smash{{\SetFigFont{12}{14.4}{\rmdefault}{\mddefault}{\updefault}{\sl fis}}}}}
\put(9930,-5946){\makebox(0,0)[lb]{\smash{{\SetFigFont{10}{12.0}{\rmdefault}{\mddefault}{\updefault}Fis}}}}
\put(9263,-6251){\makebox(0,0)[lb]{\smash{{\SetFigFont{12}{14.4}{\rmdefault}{\mddefault}{\updefault}{\sl fis}}}}}
\end{picture}%

%% file: figures/temp_profile_reduced.pstex_t
\begin{picture}(0,0)%
\includegraphics{temp_profile_reduced.pstex}%
\end{picture}%
\setlength{\unitlength}{3947sp}%
\begingroup\makeatletter\ifx\SetFigFont\undefined%
\gdef\SetFigFont#1#2#3#4#5{%
  \reset@font\fontsize{#1}{#2pt}%
  \fontfamily{#3}\fontseries{#4}\fontshape{#5}%
  \selectfont}%
\fi\endgroup%
\begin{picture}(7094,6429)(301,-5794)
\put(2036,-1621){\makebox(0,0)[lb]{\smash{{\SetFigFont{12}{14.4}{\rmdefault}{\mddefault}{\updefault}$\theta_{fis}^1$}}}}
\put(2036,-1021){\makebox(0,0)[lb]{\smash{{\SetFigFont{12}{14.4}{\rmdefault}{\mddefault}{\updefault}$\theta_{fis}^2$}}}}
\put(1586,-1321){\makebox(0,0)[lb]{\smash{{\SetFigFont{12}{14.4}{\rmdefault}{\mddefault}{\updefault}$\kappa_{fis}^1/\gamma_{fis}$}}}}
\put(2036,-721){\makebox(0,0)[lb]{\smash{{\SetFigFont{12}{14.4}{\rmdefault}{\mddefault}{\updefault}$\theta_{fis}^3$}}}}
\put(2036,-421){\makebox(0,0)[lb]{\smash{{\SetFigFont{12}{14.4}{\rmdefault}{\mddefault}{\updefault}$\theta_{fis}^4$}}}}
\put(2036,-121){\makebox(0,0)[lb]{\smash{{\SetFigFont{12}{14.4}{\rmdefault}{\mddefault}{\updefault}$\theta_{fis}^5$}}}}
\put(911,179){\makebox(0,0)[lb]{\smash{{\SetFigFont{12}{14.4}{\rmdefault}{\mddefault}{\updefault}$(\kappa_{fis}^1+\kappa_{fis}^2)/\gamma_{fis}$}}}}
\put(1736,479){\makebox(0,0)[lb]{\smash{{\SetFigFont{12}{14.4}{\rmdefault}{\mddefault}{\updefault}$\maX_{fis}$}}}}
\put(2411,-2221){\makebox(0,0)[lb]{\smash{{\SetFigFont{12}{14.4}{\rmdefault}{\mddefault}{\updefault}$D^1$}}}}
\put(3011,-2221){\makebox(0,0)[lb]{\smash{{\SetFigFont{12}{14.4}{\rmdefault}{\mddefault}{\updefault}$D^{70}$}}}}
\put(3311,-2371){\makebox(0,0)[lb]{\smash{{\SetFigFont{12}{14.4}{\rmdefault}{\mddefault}{\updefault}$D^{75}$}}}}
\put(3611,-2221){\makebox(0,0)[lb]{\smash{{\SetFigFont{12}{14.4}{\rmdefault}{\mddefault}{\updefault}$D^{78}$}}}}
\put(4211,-2221){\makebox(0,0)[lb]{\smash{{\SetFigFont{12}{14.4}{\rmdefault}{\mddefault}{\updefault}$D^{86}$}}}}
\put(4811,-2221){\makebox(0,0)[lb]{\smash{{\SetFigFont{12}{14.4}{\rmdefault}{\mddefault}{\updefault}$D^{94}$}}}}
\put(5411,-2221){\makebox(0,0)[lb]{\smash{{\SetFigFont{12}{14.4}{\rmdefault}{\mddefault}{\updefault}$D^{102}$}}}}
\put(5711,-2371){\makebox(0,0)[lb]{\smash{{\SetFigFont{12}{14.4}{\rmdefault}{\mddefault}{\updefault}$D^{105}$}}}}
\put(6011,-2221){\makebox(0,0)[lb]{\smash{{\SetFigFont{12}{14.4}{\rmdefault}{\mddefault}{\updefault}$D^{61}$}}}}
\put(6311,-2371){\makebox(0,0)[lb]{\smash{{\SetFigFont{12}{14.4}{\rmdefault}{\mddefault}{\updefault}$D^{30}$}}}}
\put(6611,-2221){\makebox(0,0)[lb]{\smash{{\SetFigFont{12}{14.4}{\rmdefault}{\mddefault}{\updefault}$D^{31}$}}}}
\put(2711,-2371){\makebox(0,0)[lb]{\smash{{\SetFigFont{12}{14.4}{\rmdefault}{\mddefault}{\updefault}$D^2$}}}}
\put(5111,-2371){\makebox(0,0)[lb]{\smash{{\SetFigFont{12}{14.4}{\rmdefault}{\mddefault}{\updefault}$D^{99}$}}}}
\put(3911,-2371){\makebox(0,0)[lb]{\smash{{\SetFigFont{12}{14.4}{\rmdefault}{\mddefault}{\updefault}$D^{83}$}}}}
\put(4511,-2371){\makebox(0,0)[lb]{\smash{{\SetFigFont{12}{14.4}{\rmdefault}{\mddefault}{\updefault}$D^{91}$}}}}
\put(2186,-1921){\makebox(0,0)[lb]{\smash{{\SetFigFont{12}{14.4}{\rmdefault}{\mddefault}{\updefault}$0$}}}}
\put(2026,-4986){\makebox(0,0)[lb]{\smash{{\SetFigFont{12}{14.4}{\rmdefault}{\mddefault}{\updefault}$\theta_{crp}^1$}}}}
\put(1576,-4686){\makebox(0,0)[lb]{\smash{{\SetFigFont{12}{14.4}{\rmdefault}{\mddefault}{\updefault}$\kappa_{crp}^1/\gamma_{crp}$}}}}
\put(901,-4386){\makebox(0,0)[lb]{\smash{{\SetFigFont{12}{14.4}{\rmdefault}{\mddefault}{\updefault}$(\kappa_{crp}^1+\kappa_{crp}^2)/\gamma_{crp}$}}}}
\put(2026,-4086){\makebox(0,0)[lb]{\smash{{\SetFigFont{12}{14.4}{\rmdefault}{\mddefault}{\updefault}$\theta_{crp}^2$}}}}
\put(2026,-3786){\makebox(0,0)[lb]{\smash{{\SetFigFont{12}{14.4}{\rmdefault}{\mddefault}{\updefault}$\theta_{crp}^3$}}}}
\put(901,-3486){\makebox(0,0)[lb]{\smash{{\SetFigFont{12}{14.4}{\rmdefault}{\mddefault}{\updefault}$(\kappa_{crp}^1+\kappa_{crp}^3)/\gamma_{crp}$}}}}
\put(301,-3186){\makebox(0,0)[lb]{\smash{{\SetFigFont{12}{14.4}{\rmdefault}{\mddefault}{\updefault}$(\kappa_{crp}^1+\kappa_{crp}^2+\kappa_{crp}^3)/\gamma_{crp}$}}}}
\put(1726,-2886){\makebox(0,0)[lb]{\smash{{\SetFigFont{12}{14.4}{\rmdefault}{\mddefault}{\updefault}$\maX_{crp}$}}}}
\put(2401,-5586){\makebox(0,0)[lb]{\smash{{\SetFigFont{12}{14.4}{\rmdefault}{\mddefault}{\updefault}$D^1$}}}}
\put(2701,-5736){\makebox(0,0)[lb]{\smash{{\SetFigFont{12}{14.4}{\rmdefault}{\mddefault}{\updefault}$D^2$}}}}
\put(3001,-5586){\makebox(0,0)[lb]{\smash{{\SetFigFont{12}{14.4}{\rmdefault}{\mddefault}{\updefault}$D^{70}$}}}}
\put(3301,-5736){\makebox(0,0)[lb]{\smash{{\SetFigFont{12}{14.4}{\rmdefault}{\mddefault}{\updefault}$D^{75}$}}}}
\put(3601,-5586){\makebox(0,0)[lb]{\smash{{\SetFigFont{12}{14.4}{\rmdefault}{\mddefault}{\updefault}$D^{78}$}}}}
\put(3901,-5736){\makebox(0,0)[lb]{\smash{{\SetFigFont{12}{14.4}{\rmdefault}{\mddefault}{\updefault}$D^{83}$}}}}
\put(4501,-5736){\makebox(0,0)[lb]{\smash{{\SetFigFont{12}{14.4}{\rmdefault}{\mddefault}{\updefault}$D^{91}$}}}}
\put(4201,-5586){\makebox(0,0)[lb]{\smash{{\SetFigFont{12}{14.4}{\rmdefault}{\mddefault}{\updefault}$D^{86}$}}}}
\put(4801,-5586){\makebox(0,0)[lb]{\smash{{\SetFigFont{12}{14.4}{\rmdefault}{\mddefault}{\updefault}$D^{94}$}}}}
\put(5101,-5736){\makebox(0,0)[lb]{\smash{{\SetFigFont{12}{14.4}{\rmdefault}{\mddefault}{\updefault}$D^{99}$}}}}
\put(5401,-5586){\makebox(0,0)[lb]{\smash{{\SetFigFont{12}{14.4}{\rmdefault}{\mddefault}{\updefault}$D^{102}$}}}}
\put(5701,-5736){\makebox(0,0)[lb]{\smash{{\SetFigFont{12}{14.4}{\rmdefault}{\mddefault}{\updefault}$D^{105}$}}}}
\put(6001,-5586){\makebox(0,0)[lb]{\smash{{\SetFigFont{12}{14.4}{\rmdefault}{\mddefault}{\updefault}$D^{61}$}}}}
\put(6301,-5736){\makebox(0,0)[lb]{\smash{{\SetFigFont{12}{14.4}{\rmdefault}{\mddefault}{\updefault}$D^{30}$}}}}
\put(6601,-5586){\makebox(0,0)[lb]{\smash{{\SetFigFont{12}{14.4}{\rmdefault}{\mddefault}{\updefault}$D^{31}$}}}}
\put(2176,-5286){\makebox(0,0)[lb]{\smash{{\SetFigFont{12}{14.4}{\rmdefault}{\mddefault}{\updefault}$0$}}}}
\end{picture}%

%% file: figures/Phase_spaceEcoli_flow.pstex_t
\begin{picture}(0,0)%
\includegraphics{Phase_spaceEcoli_flow.pstex}%
\end{picture}%
\setlength{\unitlength}{3947sp}%
\begingroup\makeatletter\ifx\SetFigFont\undefined%
\gdef\SetFigFont#1#2#3#4#5{%
  \reset@font\fontsize{#1}{#2pt}%
  \fontfamily{#3}\fontseries{#4}\fontshape{#5}%
  \selectfont}%
\fi\endgroup%
\begin{picture}(3687,3544)(3526,-4558)
\put(4651,-4261){\makebox(0,0)[lb]{\smash{{\SetFigFont{14}{16.8}{\rmdefault}{\mddefault}{\updefault}$\theta_{\cya}^2$}}}}
\put(5851,-4261){\makebox(0,0)[lb]{\smash{{\SetFigFont{14}{16.8}{\rmdefault}{\mddefault}{\updefault}$\theta_{\cya}^3$}}}}
\put(7051,-4261){\makebox(0,0)[lb]{\smash{{\SetFigFont{14}{16.8}{\rmdefault}{\mddefault}{\updefault}$\mathit{max}_{\cya}$}}}}
\put(4276,-2836){\makebox(0,0)[lb]{\smash{{\SetFigFont{14}{16.8}{\rmdefault}{\mddefault}{\updefault}$\theta^3_{\crp}$}}}}
\put(4276,-3961){\makebox(0,0)[lb]{\smash{{\SetFigFont{14}{16.8}{\rmdefault}{\mddefault}{\updefault}$\theta^2_{\crp}$}}}}
\put(4051,-1336){\makebox(0,0)[lb]{\smash{{\SetFigFont{14}{16.8}{\rmdefault}{\mddefault}{\updefault}$x_{\crp}$}}}}
\put(4051,-1636){\makebox(0,0)[lb]{\smash{{\SetFigFont{14}{16.8}{\rmdefault}{\mddefault}{\updefault}$\mathit{max}_{\crp}$}}}}
\put(3526,-2311){\makebox(0,0)[lb]{\smash{{\SetFigFont{14}{16.8}{\rmdefault}{\mddefault}{\updefault}$\frac{\kappa^1_{\crp}+\kappa^2_{\crp}+\kappa^3_{\crp}}{\gamma_{\crp}}$}}}}
\put(6226,-4486){\makebox(0,0)[lb]{\smash{{\SetFigFont{14}{16.8}{\rmdefault}{\mddefault}{\updefault}$\frac{\kappa^1_{\cya}+\kappa^2_{\cya}}{\gamma_{\cya}}$}}}}
\end{picture}%

%% file: figures/Phase_spaceEcoliCoarse.pstex_t
\begin{picture}(0,0)%
\includegraphics{Phase_spaceEcoliCoarse.pstex}%
\end{picture}%
\setlength{\unitlength}{3947sp}%
\begingroup\makeatletter\ifx\SetFigFont\undefined%
\gdef\SetFigFont#1#2#3#4#5{%
  \reset@font\fontsize{#1}{#2pt}%
  \fontfamily{#3}\fontseries{#4}\fontshape{#5}%
  \selectfont}%
\fi\endgroup%
\begin{picture}(3492,3543)(4051,-4562)
\put(4276,-2836){\makebox(0,0)[lb]{\smash{{\SetFigFont{14}{16.8}{\rmdefault}{\mddefault}{\updefault}$\theta^3_{\crp}$}}}}
\put(6376,-2836){\makebox(0,0)[lb]{\smash{{\SetFigFont{14}{16.8}{\rmdefault}{\mddefault}{\updefault}$M^{32}$}}}}
\put(6376,-3511){\makebox(0,0)[lb]{\smash{{\SetFigFont{14}{16.8}{\rmdefault}{\mddefault}{\updefault}$M^{61}$}}}}
\put(4651,-4261){\makebox(0,0)[lb]{\smash{{\SetFigFont{14}{16.8}{\rmdefault}{\mddefault}{\updefault}$\theta_{\cya}^2$}}}}
\put(5851,-4261){\makebox(0,0)[lb]{\smash{{\SetFigFont{14}{16.8}{\rmdefault}{\mddefault}{\updefault}$\theta_{\cya}^3$}}}}
\put(7051,-4261){\makebox(0,0)[lb]{\smash{{\SetFigFont{14}{16.8}{\rmdefault}{\mddefault}{\updefault}$\mathit{max}_{\cya}$}}}}
\put(4276,-3961){\makebox(0,0)[lb]{\smash{{\SetFigFont{14}{16.8}{\rmdefault}{\mddefault}{\updefault}$\theta^2_{\crp}$}}}}
\put(4051,-1636){\makebox(0,0)[lb]{\smash{{\SetFigFont{14}{16.8}{\rmdefault}{\mddefault}{\updefault}$\mathit{max}_{\crp}$}}}}
\put(5851,-2161){\makebox(0,0)[lb]{\smash{{\SetFigFont{14}{16.8}{\rmdefault}{\mddefault}{\updefault}$M^{30}$}}}}
\put(6376,-2161){\makebox(0,0)[lb]{\smash{{\SetFigFont{14}{16.8}{\rmdefault}{\mddefault}{\updefault}$M^{29}$}}}}
\end{picture}%

%% file: figures/Phase_spaceEcoliFine.pstex_t
\begin{picture}(0,0)%
\includegraphics{Phase_spaceEcoliFine.pstex}%
\end{picture}%
\setlength{\unitlength}{3947sp}%
\begingroup\makeatletter\ifx\SetFigFont\undefined%
\gdef\SetFigFont#1#2#3#4#5{%
  \reset@font\fontsize{#1}{#2pt}%
  \fontfamily{#3}\fontseries{#4}\fontshape{#5}%
  \selectfont}%
\fi\endgroup%
\begin{picture}(4672,3539)(3526,-4558)
\put(4276,-3961){\makebox(0,0)[lb]{\smash{{\SetFigFont{14}{16.8}{\rmdefault}{\mddefault}{\updefault}$\theta^2_{\crp}$}}}}
\put(4651,-4261){\makebox(0,0)[lb]{\smash{{\SetFigFont{14}{16.8}{\rmdefault}{\mddefault}{\updefault}$\theta_{\cya}^2$}}}}
\put(5851,-4261){\makebox(0,0)[lb]{\smash{{\SetFigFont{14}{16.8}{\rmdefault}{\mddefault}{\updefault}$\theta_{\cya}^3$}}}}
\put(7051,-4261){\makebox(0,0)[lb]{\smash{{\SetFigFont{14}{16.8}{\rmdefault}{\mddefault}{\updefault}$\mathit{max}_{\cya}$}}}}
\put(7876,-4336){\makebox(0,0)[lb]{\smash{{\SetFigFont{14}{16.8}{\rmdefault}{\mddefault}{\updefault}$x_{\cya}$}}}}
\put(6676,-3511){\makebox(0,0)[lb]{\smash{{\SetFigFont{14}{16.8}{\rmdefault}{\mddefault}{\updefault}$D^{61.1}$}}}}
\put(6601,-2836){\makebox(0,0)[lb]{\smash{{\SetFigFont{14}{16.8}{\rmdefault}{\mddefault}{\updefault}$D^{32.1}$}}}}
\put(4276,-2836){\makebox(0,0)[lb]{\smash{{\SetFigFont{14}{16.8}{\rmdefault}{\mddefault}{\updefault}$\theta^3_{\crp}$}}}}
\put(6226,-4486){\makebox(0,0)[lb]{\smash{{\SetFigFont{14}{16.8}{\rmdefault}{\mddefault}{\updefault}$\frac{\kappa^1_{\cya}+\kappa^2_{\cya}}{\gamma_{\cya}}$}}}}
\put(4051,-1636){\makebox(0,0)[lb]{\smash{{\SetFigFont{14}{16.8}{\rmdefault}{\mddefault}{\updefault}$\mathit{max}_{\crp}$}}}}
\put(6601,-2611){\makebox(0,0)[lb]{\smash{{\SetFigFont{14}{16.8}{\rmdefault}{\mddefault}{\updefault}$D^{29.1}$}}}}
\put(5851,-2611){\makebox(0,0)[lb]{\smash{{\SetFigFont{14}{16.8}{\rmdefault}{\mddefault}{\updefault}$D^{30.1}$}}}}
\put(5851,-2386){\makebox(0,0)[lb]{\smash{{\SetFigFont{14}{16.8}{\rmdefault}{\mddefault}{\updefault}$D^{30.2}$}}}}
\put(3526,-2311){\makebox(0,0)[lb]{\smash{{\SetFigFont{14}{16.8}{\rmdefault}{\mddefault}{\updefault}$\frac{\kappa^1_{\crp}+\kappa^2_{\crp}+\kappa^3_{\crp}}{\gamma_{\crp}}$}}}}
\end{picture}%

%% file: figures/State_graphEcoliCoarse.pstex_t
\begin{picture}(0,0)%
\includegraphics{State_graphEcoliCoarse.pstex}%
\end{picture}%
\setlength{\unitlength}{3947sp}%
\begingroup\makeatletter\ifx\SetFigFont\undefined%
\gdef\SetFigFont#1#2#3#4#5{%
  \reset@font\fontsize{#1}{#2pt}%
  \fontfamily{#3}\fontseries{#4}\fontshape{#5}%
  \selectfont}%
\fi\endgroup%
\begin{picture}(5040,912)(3826,-4117)
\put(6451,-3361){\makebox(0,0)[lb]{\smash{{\SetFigFont{12}{14.4}{\rmdefault}{\mddefault}{\updefault}$\dot{x}_{crp} ?$}}}}
\put(3826,-4036){\makebox(0,0)[lb]{\smash{{\SetFigFont{17}{20.4}{\rmdefault}{\mddefault}{\updefault}$M^{61}$}}}}
\put(5101,-4036){\makebox(0,0)[lb]{\smash{{\SetFigFont{17}{20.4}{\rmdefault}{\mddefault}{\updefault}$M^{32}$}}}}
\put(6376,-4036){\makebox(0,0)[lb]{\smash{{\SetFigFont{17}{20.4}{\rmdefault}{\mddefault}{\updefault}$M^{29}$}}}}
\put(7651,-4036){\makebox(0,0)[lb]{\smash{{\SetFigFont{17}{20.4}{\rmdefault}{\mddefault}{\updefault}$M^{30}$}}}}
\put(7726,-3361){\makebox(0,0)[lb]{\smash{{\SetFigFont{12}{14.4}{\rmdefault}{\mddefault}{\updefault}$\dot{x}_{crp} ?$}}}}
\put(3826,-3361){\makebox(0,0)[lb]{\smash{{\SetFigFont{12}{14.4}{\rmdefault}{\mddefault}{\updefault}$\dot{x}_{crp}>0$}}}}
\end{picture}%

%% file: figures/State_graphEcoliFine.pstex_t
\begin{picture}(0,0)%
\includegraphics{State_graphEcoliFine.pstex}%
\end{picture}%
\setlength{\unitlength}{3947sp}%
\begingroup\makeatletter\ifx\SetFigFont\undefined%
\gdef\SetFigFont#1#2#3#4#5{%
  \reset@font\fontsize{#1}{#2pt}%
  \fontfamily{#3}\fontseries{#4}\fontshape{#5}%
  \selectfont}%
\fi\endgroup%
\begin{picture}(4969,912)(6001,-5242)
\put(6001,-5161){\makebox(0,0)[lb]{\smash{{\SetFigFont{17}{20.4}{\rmdefault}{\mddefault}{\updefault}$D^{61.1}$}}}}
\put(6901,-5161){\makebox(0,0)[lb]{\smash{{\SetFigFont{17}{20.4}{\rmdefault}{\mddefault}{\updefault}$D^{32.1}$}}}}
\put(7801,-5161){\makebox(0,0)[lb]{\smash{{\SetFigFont{17}{20.4}{\rmdefault}{\mddefault}{\updefault}$D^{29.1}$}}}}
\put(9601,-5161){\makebox(0,0)[lb]{\smash{{\SetFigFont{17}{20.4}{\rmdefault}{\mddefault}{\updefault}$D^{30.2}$}}}}
\put(6001,-4486){\makebox(0,0)[lb]{\smash{{\SetFigFont{12}{14.4}{\rmdefault}{\mddefault}{\updefault}$\dot{x}_{\crp}>0$}}}}
\put(7726,-4486){\makebox(0,0)[lb]{\smash{{\SetFigFont{12}{14.4}{\rmdefault}{\mddefault}{\updefault}$\dot{x}_{\crp}>0$}}}}
\put(8776,-4486){\makebox(0,0)[lb]{\smash{{\SetFigFont{12}{14.4}{\rmdefault}{\mddefault}{\updefault}$\dot{x}_{\crp}>0$}}}}
\put(9601,-4486){\makebox(0,0)[lb]{\smash{{\SetFigFont{12}{14.4}{\rmdefault}{\mddefault}{\updefault}$\dot{x}_{\crp}=0$}}}}
\put(8701,-5161){\makebox(0,0)[lb]{\smash{{\SetFigFont{17}{20.4}{\rmdefault}{\mddefault}{\updefault}$D^{30.1}$}}}}
\end{picture}%